\documentclass[%
 reprint,
 amsmath,amssymb,
 aps,
]{revtex4-1}

\usepackage{graphicx}% Include figure files
\usepackage{dcolumn}% Align table columns on decimal point
\usepackage{bm}% bold math
\usepackage[mathlines]{lineno}% Enable numbering of text and display math
\usepackage[colorlinks=true,citecolor=blue,linkcolor=blue,anchorcolor=green, anchorcolor=green,urlcolor=blue]{hyperref}
\usepackage{mathrsfs}
\usepackage{xcolor}
\usepackage{float}
\usepackage{ulem}
\usepackage[utf8]{inputenc}

\begin{document}

\preprint{APS/123-QED}

\title{Thermodynamical stability of $f(R)$-AdS black holes in grand canonical ensemble}% Force line breaks with \\
%\thanks{A footnote to the article title}%

\author{Ming Zhang}
 \email{mingzhang@mail.bnu.edu.cn}

\affiliation{Department of Physics, Beijing Normal University, Beijing, 100875, China}

\date{\today}% It is always \today, today,
             %  but any date may be explicitly specified

\begin{abstract}
In this paper, $n$-dimensional ($n=4p, p$ is a positive integer) $f(R)$-AdS black holes is divided into Schwazschild-AdS (SAdS) like ones and Reissner-Nordstr\"{o}m-AdS (RN-AdS) like ones. Thermodynamical stability of them in grand canonical ensemble is investigated. Locally, we find that the RN-AdS like $f(R)$ black holes will experience either type-one or type-two phase transitions from unstable states to stable states, whereas there are only type-two phase transitions between that two states for SAdS like $f(R)$ black holes. Globally, we find that there are type-one Hawking-Page like phase transitions between thermal AdS state and  $f(R)$ black holes. Using thermodynamical geometry method, we find that thermodynamical extrinsic curvature can only provide accurate stability information near type-two (not type-one) phase transition points for $f(R)$ black holes in grand canonical ensemble.
%\begin{description}
%\item[PACS numbers]
%04.70.Dy, 04.70.-s
%\end{description}
\end{abstract}

%\pacs{04.70.Dy, 04.70.-s}% PACS, the Physics and Astronomy
                             % Classification Scheme.

\maketitle

%\tableofcontents

\section{Introduction}%\label{Introduction}
The investigation of anti-de Sitter black holes is mainly derived from string theory, which incorporates black holes as soliton D-branes, or branes as higher-dimensional progenitors \cite{Maldacena:1997re,Witten:1998qj}. Black hole thermodynamics in AdS spacetime has attracted lots of attentions \cite{Chamblin:1999tk,Kastor:2009wy,Kubiznak:2012wp,Johnson:2014yja,Wei:2015iwa,Pourhassan:2015cga,Zhang:2018nep,Kubiznak:2016qmn,Banerjee:2016nse,Lee:2017ero,Wei:2017mwc,Chabab:2018lzf}. Hawking and Page found that there is a phase transition between SAdS black hole and thermal gas in AdS state \cite{Hawking:1982dh}. Due to AdS/CFT correspondence, which relates quantum gravity in AdS spacetime with a conformal field theory at the lower dimensional boundary of AdS, Hawking-Page phase transition is also interpreted as confinement/deconfinement phase transition of the dual gauge field theory \cite{Witten:1998zw}. Thermodynamics of RN-AdS black hole in both canonical ensemble and grand canonical ensemble is studied and a first-order phase transition between small and large black holes was found \cite{Chamblin:1999tk,Kastor:2009wy}. It is found that phase transition of AdS black hole in canonical ensemble is similar to Van der Waals liquid/gas transitions \cite{Kubiznak:2012wp}. 

$f(R)$ gravity, where the Lagrangian density is an arbitrary function of curvature scalar $R$, is a valuable modification to general relativity \cite{DeFelice:2010aj}. The study of $f(R)$ black holes is meaningful at least for two reasons. Firstly, the thermodynamics of $f(R)$-AdS black holes can give rise to  an interesting description and a better understanding of AdS/CFT correspondence. Secondly, the applications of $f(R)$ gravity in cosmology are substantial. For example, the model with $f(R)=\alpha R^{2}(\alpha>0)$ can lead to accelerated expansion of the Universe \cite{Starobinsky:1980te}. 

Although it was shown that the laws of black holes are universal, the properties of black holes are dimension dependent \cite{Altamirano:2013ane} as well as ensemble dependent \cite{Dolan:2013yca}. As not much is known about the thermodynamical stability of $f(R)$-AdS black holes in grand canonical ensemble, which is defined by coupling the system to energy and charge reserviors at fixed temperature $T$ and potential $\Phi$, we aim to fill this gap. It is worth mentioning that related meaningful investigations on thermodynamical stability of AdS black holes can be seen in Refs. \cite{Carter:2005uw,Cai:2014znn,Fernando:2016sps,Fernando:2016qhq,Li:2016wzx,Dehghani:2017zkm,Dehghani:2017ckb}.

The layout of this paper is as follows. In Sec. \ref{review}, we will briefly present an $n$-dimensional charged black hole solution with constant curvature in $f(R)$ gravity. In Sec. \ref{stability}, we will investigate the thermodynamical stability of $f(R)$-AdS black holes locally and globally. In Sec. \ref{curvature}, thermodynamical extrinsic curvature are used to study the thermodynamical stability of the  $f(R)$-AdS black holes.  Conclusions will be drawn in Sec. \ref{curvature}.

\section{Review of $n$-dimensional $f(R)$ black holes}\label{review}
Specific range of the advanced physics theories require existence of the higher dimensional gravity. The action for $f(R)$ gravity with Maxwell term in $n$-dimensional  spacetime reads
\begin{equation}
I=\frac{1}{16\pi}\int d^{n}x \sqrt{-g}\left[R+f(R)-\left(F_{\mu\nu}F^{\mu\nu}\right)^{p}\right],
\end{equation}
where $F_{\mu\nu}=\partial_{\mu}A_{\nu}-\partial_{\nu}A_{\mu}$ is the electromagnetic field tensor and the integer $p$ is the power of traceless conformally invariant Maxwell field, $R$ is the scalar curvature, $f(R)$ is an arbitrary function of $R$. A static spherically symmetric solution of the corresponding equations of motion has been obtained in Ref. \cite{Sheykhi:2012zz} for a constant scalar curvature  $R=R_{0}$ as
\begin{equation}
ds^{2}=-N(r)(r)dt^{2}+\frac{dr^{2}}{N(r)}+r^{2}d\Omega_{n-2}^{2},\label{metric}
\end{equation}
where
\begin{equation}\label{thirdterm}
N(r)=1-\frac{2m}{r^{n-3}}+\frac{q^{2}}{r^{n-2}}\cdot \frac{\left(-2q^{2}\right)^{\frac{n-4}{4}}}{b}-\frac{R_{0}}{n(n-1)}r^{2},
\end{equation}
\begin{equation}
b=1+f^{\prime}(R_{0}),~~ f^{\prime}(R)\equiv \frac{df(R)}{dR},\nonumber
\end{equation}
\begin{equation}
  m=\frac{8\pi M}{(n-2)b\Omega_{n-2}},~~~q=\left(\frac{16\sqrt{b}Q}{n(-2)^{\frac{n-4}{4}}\Omega_{n-2}}\right)^{\frac{2}{n-1}}.\nonumber
\end{equation}
Here $M, Q$ and $d\Omega^{2}_{n-2}$ denote the ADM mass, electric charge, and the metric of a unit ($n-2$)-dimensional sphere, respectively. $\Omega_{n-2}=\frac{2\pi^{(n-1)/2}}{\Gamma\left( (n-1)/2\right)}$ is the volume of an ($n-2$)-dimensional sphere. The spacetime described by this metric becomes asymptotically AdS when $R_{0}=-n(n-1)/l^{2}<0$. We will study $f(R)$-AdS black holes hereinafter. The mass, Hawking temperature, entropy, electric potential can be derived as \cite{Sheykhi:2012zz}
\begin{equation}
\begin{aligned}
M=&\frac{(2-n)  \left(2^{n/4} r^2 \left(-q^2\right)^{n/4}-2 b r^n\right)\Omega_{n-2}}{32 \pi  (n-1) n r^3}\\&+\frac{2(2-n) \Omega_{n-2} b R_0 r^{n+2}}{32 \pi  r^3},
\end{aligned}
\end{equation}
\begin{equation}
T=\frac{2^{\frac{n}{4}-3} \left(-q^2\right)^{n/4} r^{1-n}}{\pi  b}-\frac{R_0 r}{4 \pi  n}+\frac{n}{4 \pi  r}-\frac{3}{4 \pi  r},\label{temp n dim}
\end{equation}
\begin{equation}
S=\frac{b\Omega_{n-2}r_{+}^{n-2}}{4},\label{entropy n dim}
\end{equation}
\begin{equation}
\Phi=\frac{q\sqrt{b}}{r_{+}},
\end{equation}
 with $r_{+}$ denotes the outer event horizon. The temperature in Eq. (\ref{temp n dim}) is obtained by 
\begin{equation}
T=\frac{1}{4\pi}\left(\frac{dN(r)}{dr}\right)_{r=r_+},
\end{equation}
and the electric potential is obtained by Noether charge method.
The first law of thermodynamics is
\begin{equation}
dM=TdS+\Phi dQ,
\end{equation}
is indeed satisfied as one can check that
\begin{equation}
 T=\left(\frac{\partial M}{\partial S}\right)_{Q},~\Phi=\left(\frac{\partial M}{\partial Q}\right)_{S}.
\end{equation}

As an extension of Maxwell action in arbitrary dimensions that is traceless and possesses the conformal invariance, one should notice that  $n=4p$, i.e., $n=4,8,12,\cdots$\cite{Sheykhi:2012zz}. The real solutions only exist for specific dimensions which are multiples of four. Besides, one can see that the geometry of the $f(R)$ black hole is similar to the SAdS black hole with a single horizon when $p$ is even, because the second as well as the third terms of the right hand side of Eq. (\ref{thirdterm}) become negative. In contrast, the geometry is similar to RN-AdS black hole with two horizons when $p$ is odd, as the third term of the right hand side of Eq. (\ref{thirdterm}) becomes positive, just like the corresponding RN-AdS counterpart.  As a case study, we have plotted diagrams of mass $M$ vs. horizon radius $r_{+}$ in terms of electric potential $\Phi$ in Fig. \ref{figm} for $n=4$ and $n=8$, which is shown in Appendix. We see from the figures that the mass of the RN-AdS like black hole is positive all the time; however, there exists a minimal horizon radius $r_{M}$ where the mass vanishes for the SAdS like black hole. When $r_{+}<r_{M}$, the black hole with negative mass is unphysical.

\section{Local and global stability of $f(R)$-AdS black holes}\label{stability}
\subsection{Behavior of temperature}
The Hawking temperature can be reexpressed as a function in terms of electric potential as
\begin{equation}\label{temps}
T=\frac{2^{\frac{n}{4}-3} r_{+}^{1-n} \left(-\frac{r_{+}^2 \Phi ^2}{b}\right)^{n/4}}{\pi  b}-\frac{r_{+} R_0}{4 \pi  n}+\frac{n-3}{4 \pi  r_{+}}.
\end{equation}
We can analyze the monotonicity of the temperature by its first-order derivative of $r_{+}$ which reads
\begin{equation}
\begin{aligned}
\left(\frac{\partial T}{\partial r_{+}}\right)_{\Phi}=&\frac{\left(2^{\frac{n}{4}-3}-2^{\frac{n}{4}-4} n\right) r_{+}^{-n} \left(-\frac{r_{+}^2 \Phi ^2}{b}\right)^{n/4}}{\pi  b}\\&-\frac{n-3}{4 \pi  r_{+}^2}-\frac{R_0}{4 \pi  n}.
\end{aligned}
\end{equation}

We here analyze cases of $n=4$ and $n=8$. When $n=4$, the equation
\begin{equation}
\left(\frac{\partial T}{\partial r_{+}}\right)_{\Phi}=\frac{4 \left(\Phi ^2-b^2\right)-b^2 r^2 R_0}{16 \pi  b^2 r^2}=0\nonumber
\end{equation}
exists two real roots if $\Phi<b$, such that there is a minimal value for temperature; otherwise, the temperature is monotonically increasing with $r_{+}$. When $n=8$, the equation
\begin{equation}
\left(\frac{\partial T}{\partial r_{+}}\right)_{\Phi}=\frac{-b^3 r^4 R_0-40 b^3 r^2-48 \Phi ^4}{32 \pi  b^3 r^4}=0\nonumber
\end{equation}
exists two real roots all the time, so there must exists one minimal temperature. In fact, we can get the result by analyzing the Eq. (\ref{temps}) directly. As we can see, when $n=8$, all coefficients of $r_+$ are positive. When $r_+$ is small, the first and third term play the leading role, leading to the decreasing of the temperature with the increasing horizon; when $r_+$ is relatively large, the second term becomes to make effect, which leads to the increasing temperature following with the increasing horizon.

We have shown figures of temperature $T$ vs. horizon radius $r_{+}$ in Fig. \ref{figt} for $n=4$ and $n=8$, corresponding to RN-AdS like and SAdS like $f(R)$ black holes, respectively. It should be noticed that the temperature is monotonically increasing with $r_{+}$ for 4-dimensional RN-AdS like $f(R)$ black hole with $\Phi>b$. In this context, there exists a minimal horizon radius corresponding to an extremal black hole where the Hawking temperature vanishes.

\subsection{Specific heat capacity, electrical permittivity and local stability}

Mentioning the stability of a thermodynamical system, it is common to consider small fluctuations of state functions around equilibrium, and since first-order terms vanish, the stability is only a statement about second-order variations. In the grand canonical ensemble, an equivalent manner of studying the local stability can be done by analyzing the heat capacity $C_{\Phi}$ at constant electric potential and the electrical permittivity $\epsilon_{T}$ at constant temperature. The black hole will be thermally or electrically unstable under fluctuations if the heat capacity or electrical permittivity is negative. The local stability of a system will be ensured only if there exists a range of horizon radius for which the quantities $C_{\Phi}$ and $\epsilon_{T}$ are both positive. In another way, the nonstable black holes may undergo a phase transition to be stabilized. The phase transition points are those which make the heat capacity and electrical permittivity vanish or diverge. Phase transitions at the vanishing points (roots of heat capacity) belong to type-one type. Parallelly, phase transitions at divergent points belong to type-two type.

The specific heat at constant electric potential for $f(R)$ black holes can be derived as
\begin{widetext}
\begin{equation}
C_{\Phi}\equiv T\left(\frac{\partial S}{\partial T}\right)_{\Phi}=-\frac{b (n-2) \Omega_{n-2}  r_{+}^{n-2} \left[n \left(2 b (n-3) r_{+}^n+2^{n/4} r_{+}^2 \left(-\frac{r_{+}^2 \Phi ^2}{b}\right)^{n/4}\right)-2 b R_0 r_{+}^{n+2}\right]}{8 b R_0 r_{+}^{n+2}+2n \left(4 b (n-3) r_{+}^n+2^{n/4} (n-2) r_{+}^2 \left(-\frac{r_{+}^2 \Phi ^2}{b}\right)^{n/4}\right)}.
\end{equation}
\end{widetext}
The electrical permittivity at constant temperature is
\begin{equation}\label{epsilon}
\epsilon_{T}\equiv \left(\frac{\partial Q}{\partial \Phi}\right)_{T}=\left(\frac{\partial\Phi}{\partial r_{+}}\right)^{-1}_{T}\left(\frac{\partial Q}{\partial r_{+}}\right)_{T}.
\end{equation}
To complete the calculation, we rewrite each factor of the above formula as
\begin{equation}
\left(\frac{\partial Q}{\partial r_{+}}\right)_{T}=-\left(\frac{\partial T}{\partial Q}\right)^{-1}_{r_{+}}\left(\frac{\partial T}{\partial r_{+}}\right)_{Q},
\end{equation}
\begin{equation}\label{epsilontwo}
\left(\frac{\partial \Phi}{\partial r_{+}}\right)_{T}=-\left(\frac{\partial T}{\partial \Phi}\right)^{-1}_{r_{+}}\left(\frac{\partial T}{\partial r_{+}}\right)_{\Phi}.
\end{equation}
Combining Eqs. (\ref{epsilon})-(\ref{epsilontwo}), the electrical permittivity can be calculated as
\begin{widetext}
\begin{equation}
\epsilon_{T}=\frac{(-1)^{\frac{n+4}{4}} (n-2) n \Omega_{n-2}  \left(\frac{r_{+} \Phi }{\sqrt{b}}\right)^{n/2} \left[2 b r_{+}^n \left((n-3) n+r_{+}^2 R_0\right)+2^{n/4} (n-1) n r_{+}^2 \left(-\left(q^{\frac{n}{2}-1}\right)^{\frac{4}{n-2}}\right)^{n/4}\right]}{32 \pi  r_{+} \Phi ^2 \left[(n-2) n r_{+}^2 \left(-\frac{r_{+}^2 \Phi ^2}{b}\right)^{n/4}+b 2^{2-\frac{n}{4}} r_{+}^n \left((n-3) n+r_{+}^2 R_0\right)\right]}.
\end{equation}
\end{widetext}

As a case study, we plot figures of specific heat capacities vs. horizon radius in Fig. \ref{figc} and electrical permittivity vs. horizon radius in Fig. \ref{fige} for 4-dimensional RN-AdS like and 8-dimensional SAdS like $f(R)$ black holes. 

Behaviors of specific heat and electrical permittivity, as can be seen in the figures, depend on the electrical potential for 4-dimensional RN-AdS like $f(R)$ black holes. Firstly, we inspect the local stability in condition of  relatively smaller electrical potential $(\Phi=1<b)$.  In the range of $r_{+}<r_{T4}=0.430$($r_{T4}$ is the turning point in Fig. \ref{figt}), both the heat capacity and electrical permittivity of the physical black hole (with positive mass and temperature) are negative, which manifests that the black holes are locally unstable. On the other hand, if $r_{+}>r_{T4}$,  both the heat capacity and electrical permittivity are positive. It means that the physical black holes with the horizon radius in this range are thermodynamically stable. One can see that the denominators of both the heat capacity and electrical permittivity vanish at $r_{+}=r_{T4}$. Specific heat $C_{\Phi}$ and electrical permittivity $\epsilon_{T}$ diverges at $r_{+}=r_{T4}$. Besides, the signs of $C_{\Phi}$ and $\epsilon_{T}$ are flipped at $r_{+}=r_{T4}$.  Radical change in the thermodynamical local stability of the system takes place at the divergent point. 

The $f(R)$ black holes undergo type-two phase transitions. Then we observe the local stability in condition of relatively larger electrical potential $(\Phi=3>b)$. We see from the figures that the electrical permittivity is positive all the time, telling us that the black hole is electrically stable all the time. However, we see that if $r_{+}<r_{T4}^{'}$, the heat capacity is negative, corresponding to a locally unstable black hole; on the other hand, if $r_{+}>r_{T4}^{'}$, the heat capacity is positive, corresponding to a locally stable black hole. The black hole experience a phase transition at the point $r_{+}=r_{T4}^{'}$.  Different from the case previously mentioned, the phase transition here belongs to type-one type, as the heat capacity vanishes rather than diverges at the phase transition point.

For 8-dimensional Schwarzschild-AdS like $f(R)$ black holes, the behaviors of specific heat and electric permittivity is similar to that of 4-dimensional RN-AdS like $f(R)$ black holes with relatively smaller electrical potential. Both the heat capacity and electrical permittivity of the unphysical black hole (with negative mass) are negative if $r_{+}<r_{T8}$, indicating that the black hole is locally unstable. Both the heat capacity and electrical permittivity of the black hole are positive if $r_{T8}<r_{+}<r_{\epsilon}$, where $r_{\epsilon}$ is defined as the value which makes the electrical  permittivity vanish, i.e., $\epsilon (r_{\epsilon})=0$, indicating that the black hole is locally stable. It should be noticed that just the black hole with horizon radius $r_{T8}<r_{M}<r_{+}<r_{\epsilon}$ is physical. In fact, whatever the value of electric potential $\Phi$ is, we find that the SAdS like  black hole will experience a type-two phase transition from unstable state to stable state with increasing horizon radius $r_{+}$.

\subsection{Gibbs free energy and global stability}
The Gibbs free energy is an appropriate state function to compare configurations in the grand canonical ensemble. Global stability of black holes can be understood by studying the corresponding Gibbs free energy.  The Gibbs free energy for the $f(R)$ black hole in the grand canonical ensemble is 
\begin{widetext}
\begin{equation}
%\begin{aligned}
G=M-TS-Q\Phi=\frac{  2 \Omega_{n-2} b R_0 r_{+}^{n+2}+(n-1)  n \Omega_{n-2} \left[2 b r_{+}^n-2^{n/4} r_{+}^2 \left((n-1) \left(-\frac{r_{+}^2 \Phi ^2}{b}\right)^{n/4}-e^{\frac{i \pi  n}{4}} n \left(\frac{r_{+} \Phi }{\sqrt{b}}\right)^{n/2}\right)\right]}{32 \pi  (n-1) n r_{+}^3}.
%\end{aligned}
\end{equation}
\end{widetext}

As a case study, we show figures of Gibbs free energy vs. temperature in Fig. \ref{figg} for 4-dimensional RN-AdS like and 8-dimensional SAdS like $f(R)$ black holes. 

Firstly, we study the global stability for 4-dimensional RN-AdS like $f(R)$ black holes in condition of relatively smaller electrical potential ($\Phi=1<b$).   When $T<T_{min}$, no black hole exists, the thermal AdS state is thermodynamically preferred. When $T>T_{min}$, there are two branches of black holes, meeting at a cusp like point. The upper branch corresponds to a small black hole with negative specific heat, which are thermodynamically unstable. The lower branch corresponds to a larger black hole, which are thermodynamically stable. However, beyond the points $T=T_{H}$, the Gibbs free energy of both the branches are positive. Hence the thermal AdS states are globally preferred. For the temperature $T>T_{L}$, the larger black hole are thermodynamically preferred over thermal AdS state since it has smaller free energy than thermal AdS state. It is obvious that the type-one Hawking-Page like phase transition between large black hole and thermal AdS space occurs at $T=T_{H}$.

In comparison, the behavior of Gibbs free energy is drastically different for 4-dimensional RN-AdS like $f(R)$ black holes with relatviely larger electrical potential ($\Phi=3>b$). The Gibbs free energy is a negative decreasing function, hence the black hole configurations are more likely than the thermal AdS background.

For 8-dimensional SAdS like $f(R)$ black holes, the behaviors of Gibbs free energy is similar to that of 4-dimensional RN-AdS like ones with relatively smaller electrical potential ($\Phi<b$). For the temperature $T<T_{H}^{'}$, the thermal AdS states are globally preferred; for the temperature $T>T_{H}^{'}$, the larger black hole branch is globally preferred. A Hawking-Page like phase transition occurs at $T=T_{H}^{\prime}$.

\section{thermodynamical extrinsic curvature and stability of $f(R)$-AdS black holes}\label{curvature}
We now study thermodynamics for $f(R)$ black holes in the grand canonical ensemble using Quevedo geometry \cite{Quevedo:2006xk}. After defining a potential $\bar{M}=M-Q\Phi$, we then introduce a 5-dimensional phase space $\mathcal{T}$ with coordinates $\bar{M},S,T,Q,\Phi$, a contact 1-form $\Theta=d\bar{M}-TdS+Qd\Phi$
 which satisfies the condition  $\Theta\wedge (d\Theta)^{3}\neq 0$, and a Legendre invariant metric
\begin{equation}
G=(d\bar{M}-TdS+Qd\Phi)^{2}+TS(-dTdS-dQd\Phi),
\end{equation}
the introduced metric
\begin{equation}
g=\varphi^{*}(G)=S\frac{\partial \bar{M}}{\partial S}\left(-\frac{\partial^{2}\bar{M}}{\partial S^{2}}dS^{2}-\frac{\partial^{2}\bar{M}}{\partial \Phi^{2}}d\Phi^{2}\right)\nonumber
\end{equation}
determines all the geometric properties of the equilibrium space $\mathcal{E}$. In the above expression, we have used the Euler identity to simplify the form of the conformal factor.

It is a matter of calculation to obtain the intrinsic Riemann scalar curvature
\begin{widetext}
\begin{equation}
\begin{aligned}
R=&\frac{\mathcal{R}(r_{+},\Phi)}{(n-2) \Omega_{n-2} ^2 \left[2 b n (n-3) r_{+}^n+2^{n/4} n r_{+}^2 \left(-\frac{r_{+}^2 \Phi ^2}{b}\right)^{n/4}-2 b R_0 r_{+}^{n+2}\right]^3}\\& \times \frac{1}{\left[4 b R_0 r_{+}^{n+2}+4 b n (n-3) r_{+}^n+2^{n/4} n (n-2) r_{+}^2 \left(-\frac{r_{+}^2 \Phi ^2}{b}\right)^{n/4}\right]^2},
\end{aligned}
\end{equation}
\end{widetext}
where $\mathcal{R}(r_{+},\Phi)$  is a polynomial without any singularity.

The thermodynamical extrinsic curvature is raised in Ref. \cite{Mansoori:2016jer}  which shows that intrinsic curvature scalar can only reflect phase transition points of the black hole whereas extrinsic curvature scalar can not only reflect the phase transition points, but also the thermodynamic stability of the black hole, as the extrinsic curvature scalar shares the same signs near the phase transition points with the specific heat, which does not happen for the intrinsic curvature scalar \cite{Zhang:2018djl,Wang:2018civ}. We have noticed that the result about the comparative advantage of the extrinsic curvature scalar on the judgement of the thermodynamic stability near the phase transition point is obtained only in the canonical ensemble, where the electric charge of the black hole is keeped unchanged. What about the result when we change to the grand canonical ensemble, where the electric potential is constant? There is however no investigation on this subject so far. So we will clarify this for $f(R)$-AdS black holes.  We here restrict ourselves to live on a constant $\Phi$ hypersurface with a normalized normal covector 
\begin{equation}
n_{\Phi}=\frac{-\Psi_{,\phi}}{\sqrt{g^{ab}\Psi_{,a}\Psi_{,b}}}=\frac{-1}{\sqrt{|g^{\Phi\Phi}|}},
\end{equation}
where $\Psi=\Phi-$Const.
Then the extrinsic curvature is 
\begin{equation}
K=h^{ab}K_{ab}=n^{\alpha}_{;\alpha}=\frac{1}{\sqrt{g}}\partial_{\alpha}(\sqrt{g}n^{\alpha})=\partial_{\alpha}n^{\alpha}+\frac{1}{2g}n^{\alpha}\partial_{\alpha}g,
\end{equation}
where $g=\text{det}(g_{ab}),$ $h^{ab}$ is the corresponding induced metric and
\begin{equation}
n^{\alpha}=(n^{S},n^{\Phi})=(g^{S\Phi},g^{\Phi\Phi})n_{\Phi}=(0,n^{\Phi}).
\end{equation}
Then we obtain the extrinsic curvature of $f(R)$ black holes
\begin{widetext}
\begin{equation}
K=\frac{\mathcal{K}(r_{+},\Phi)}{r_{+}^2 \Phi ^3 \left[4 b R_0 r_{+}^{n+2}+n \left(4 b (n-3) r_{+}^n+2^{n/4} (n-2) r_{+}^2 \left(-\frac{r_{+}^2 \Phi ^2}{b}\right)^{n/4}\right)\right]},
\end{equation}
where 
\begin{equation*}
\begin{aligned}
\mathcal{K}(r_{+},\Phi)=&\pi  (-1)^{\frac{2n+4}{4}} 2^{\frac{n+9}{2}} (n-2) n^3 \Omega_{n-2} ^2 \frac{r\phi}{b}    \\&\times \left(\frac{(-2)^{-\frac{n}{4}} r_{+}^4 \Phi ^2 \left(\frac{r_{+} \Phi }{\sqrt{b}}\right)^{-\frac{n}{2}}}{\left|(n-2) \Omega_{n-2} ^2 \left(n \left(2 b (n-3) r_{+}^n+2^{n/4} r_{+}^2 \left(-\frac{r_{+}^2 \Phi ^2}{b}\right)^{n/4}\right)-2 b R_0 r_{+}^{n+2}\right)\right|}\right)^{3/2}\\&\times \left(\left[b (n-3) n r_{+}^n+2^{n/4} (n-2) r_{+}^2 \left(-\frac{r_{+}^2 \Phi ^2}{b}\right)^{n/4}\right]-b (n-4) R_0 r_{+}^{n+2}\right).
\end{aligned}
\end{equation*}
For 4-dimensional RN-AdS like $f(R)$ black holes, the thermodynamical extrinsic curvature reads
\begin{equation}
K=-\frac{256 \sqrt{2} r_{+}^5 \Phi  \sqrt{b} \left(b^2-\Phi ^2\right)}{\left|\frac{8 r_{+}^4 \left(b^2-\Phi ^2\right)}{b}-2 b r_{+}^6 R_0\right|^{3/2} \left(b^3 r_{+}^2 R_0+4 b^3-4 b \Phi ^2\right)},
\end{equation}
for 8-dimensional SAdS like $f(R)$ black holes, the thermodynamical extrinsic curvature is
\begin{equation}
K=\frac{640 \sqrt{3} r_{+}^6 \Phi ^2 \left(b^3 r_{+}^4 R_0-80 b^3 r_{+}^2-48 \Phi ^4\right)}{\pi ^2 b \left|\frac{32 r_{+}^6 \Phi ^4}{b^2}-2 b r_{+}^{10} R_0+80 b r_{+}^8\right|^{3/2} \left(b^3 r_{+}^4 R_0+40 b^3 r_{+}^2+48 \Phi ^4\right)}.
\end{equation}
\end{widetext}
As a case study, we show figures of intrinsic Riemann scalar curvature and extrinsic curvature vs. horizon radii in Fig. \ref{figr} and Fig. \ref{figk}. One can see from the figures that both the intrinsic curvature and extrinsic curvature are divergent just at the points where type-one and type-two phase transition take place. 

We see that in our example, on the one hand, for 4-dimensional RN-AdS like $f(R)$ black holes with relatively smaller electric potential ($\Phi=1<b$), or the 8-dimensional SAdS like $f(R)$ black holes with any electric potential, the specific heats are divergent just at the type-two phase transition points. On the other hand, for 4-dimensional RN-AdS like $f(R)$ black holes with relatively larger electric potential ($\Phi=3>b$), the specific heat is vanishing at the type-one phase transition point. Comparing the figures, we find that the intrinsic curvature cannot reflect any information of stability near the phase transition points.  In contrast, though the extrinsic curvature  does not possess the same sign as the specific heat around the type-one phase transition point, it has the same sign as the specific heat capacity around the type-two phase transition points. Namely, the extrinsic curvatures can give us information of local stability for $f(R)$ black holes which undergo local type-two phase transitions in the grand canonical ensemble.

\section{Conclusion}

$n$-dimensional $f(R)$-AdS black holes can be divided into RN-AdS like ones (such as n=4) and SAdS like ones (such as n=8). In this paper, we have studied the local and global thermodynamical stability for $n-$dimensional $f(R)$-AdS black holes in the grand canonical ensemble. Thermodynamical quantities such as specific heat capacities at constant electric potential, electric permittivity at constant temperature and Gibbs free energy have been calculated and analyzed.

For the RN-AdS like $f(R)$ black holes, the value of electric potential have great effects on its thermodynamical stability. There exists a critical value $\Phi_{*}$ for electric potential $\Phi$ (For example, when n=4, $\Phi_{*}=b$), when $\Phi<\Phi_{*}$, the black holes possess minimal temperature, locally experience type-two phase transitions from unstable state to stable state and globally undergo Hawking-Page like phase transitions from thermal AdS states to larger black holes. When $\Phi>\Phi_{*}$, the temperature of the black holes will increase monotonically with increasing horizon radius. Correspondingly, the black holes locally experience type-one phase transitions from unstable state to stable state locally and the black hole configurations are thermodynamically preferred rather than the thermal AdS states all the time globally.

For the SAdS like $f(R)$ black holes, the value of electric potential has no effects on its thermodynamical stability. On the one hand, the black holes will change from locally unstable states to stable states through a type-two phase transition, for any values of electrical potential. On the other hand, the Gibbs free energy is  a negative decreasing function which results in the black hole configurations being globally more likely than the thermal AdS background.

Employing thermodynamical geometry method, we also find that the intrinsic Riemann scalar curvature cannot give any information of stability for $f(R)$ black holes in grand canonical ensemble while the extrinsic curvature can give exact information of stability near the type-two phase transition points ( not the type-one phase transition points). This is a necessary supplementary of Ref. \cite{Mansoori:2016jer}, where the extrinsic curvature was shown to give exact information of stability near phase transition points in the canonical ensemble.
 
Our work provide a relatively comprehensive analyze of thermodynamical stability for $f(R)$-AdS black holes in the grand canonical ensemble and clarify the thermodynamical extrinsic curvature's function on judgement of thermodynamical stability, that is, thermodynamical extrinsic curvature can only provide information of stability of AdS black holes near the type-two phase transition points.

\section*{Acknowledgments}
This work is supported by the National Natural Science Foundation of China (Grant No. 11235003).

\begin{widetext}
\section*{Appendix: Figures}

 \begin{figure}[H]
 \centering
        \includegraphics[angle=0,width=0.28\textwidth]{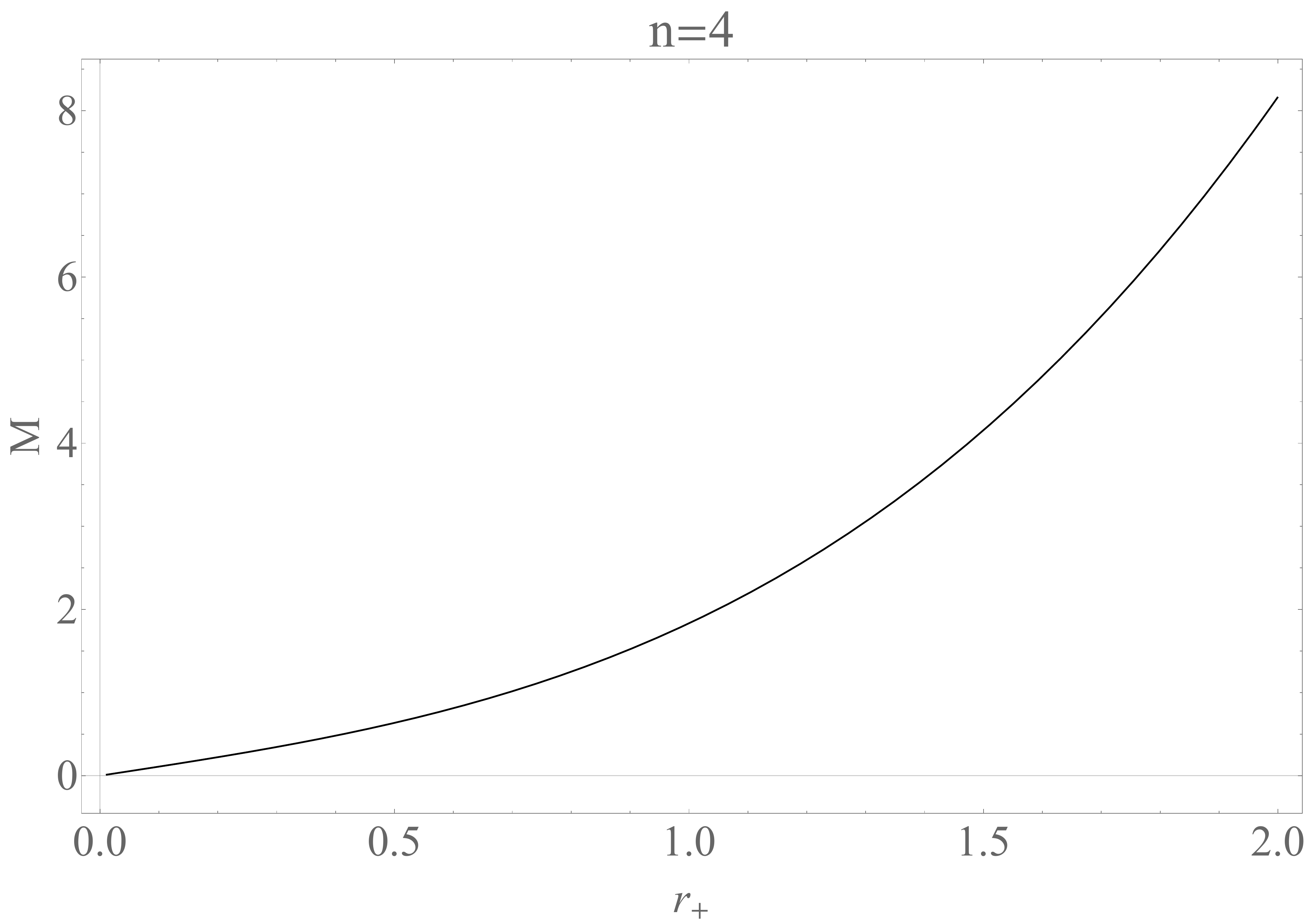}
       \includegraphics[angle=0,width=0.29\textwidth]{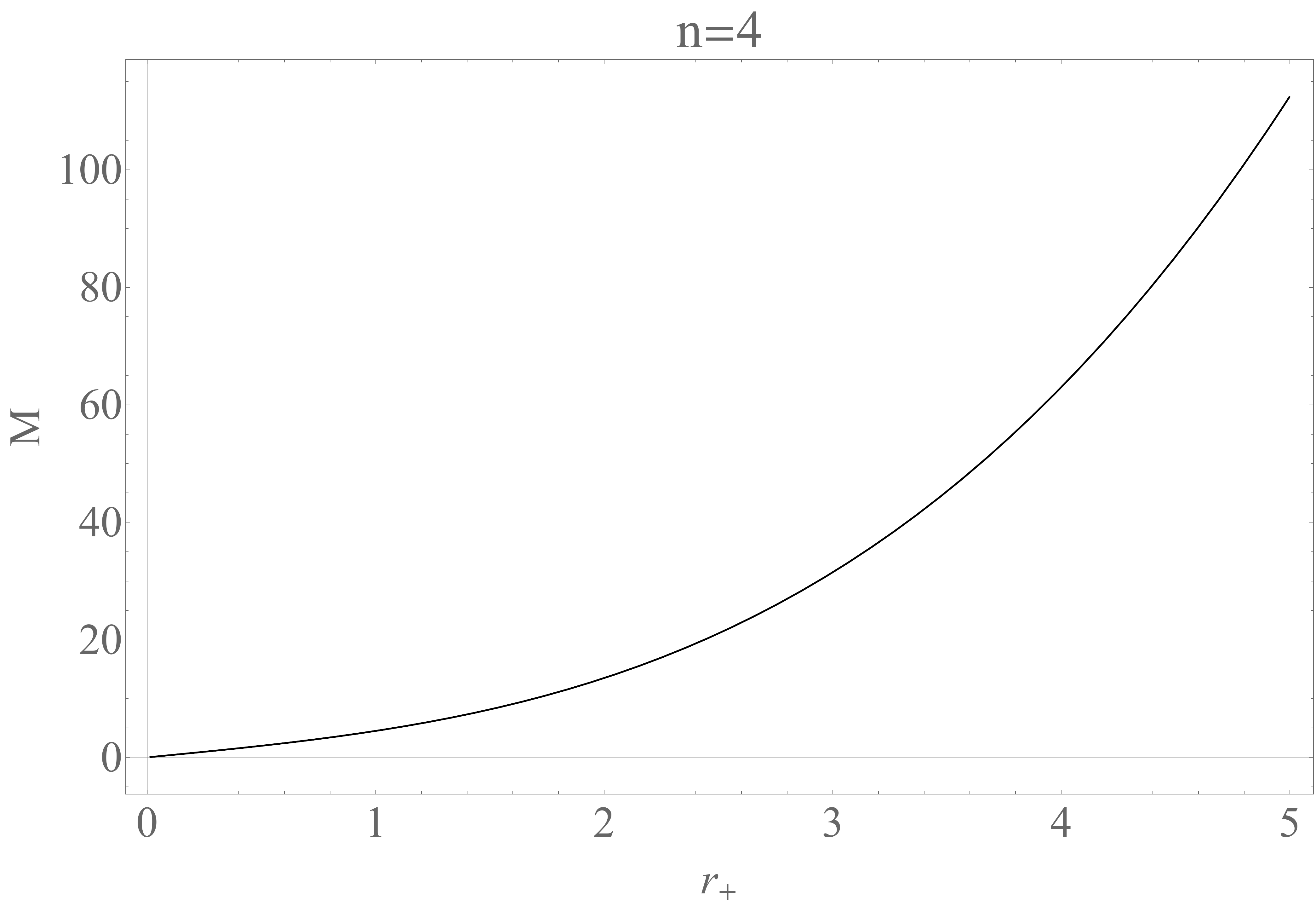}
        \includegraphics[angle=0,width=0.3\textwidth]{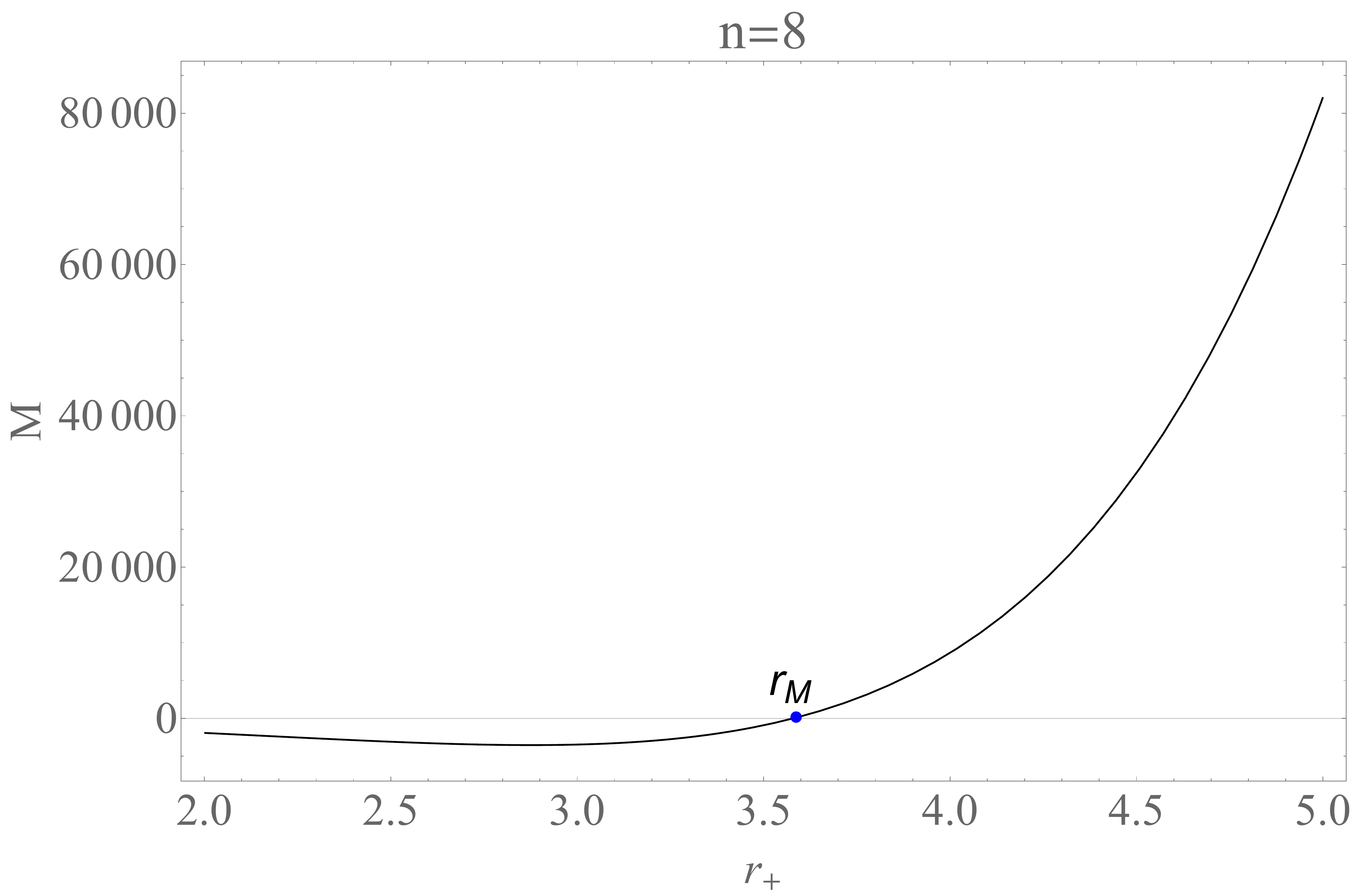}
	\caption{Mass $M$ as a function of horizon radius $r_{+}$ for $R_{0}=-12, b=1.5.$ Left: $n=4, \Phi=1$. Middle: $n=4, \Phi=3$. Right: $n=8, \Phi=3, r_{M}=3.580$.}
  \label{figm}
\end{figure}

 \begin{figure}[H]
 \centering
        \includegraphics[angle=0,width=0.29\textwidth]{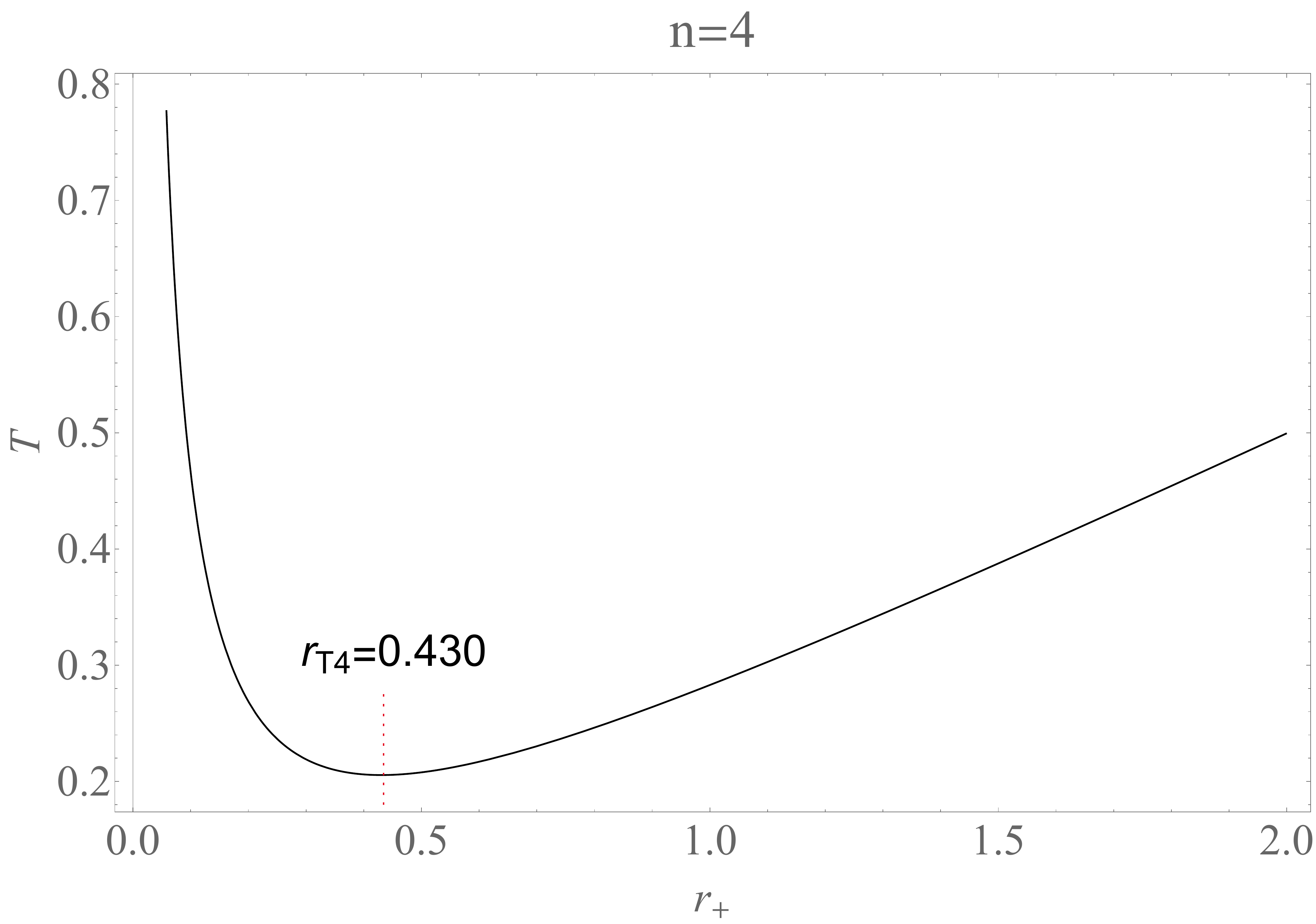}
       \includegraphics[angle=0,width=0.295\textwidth]{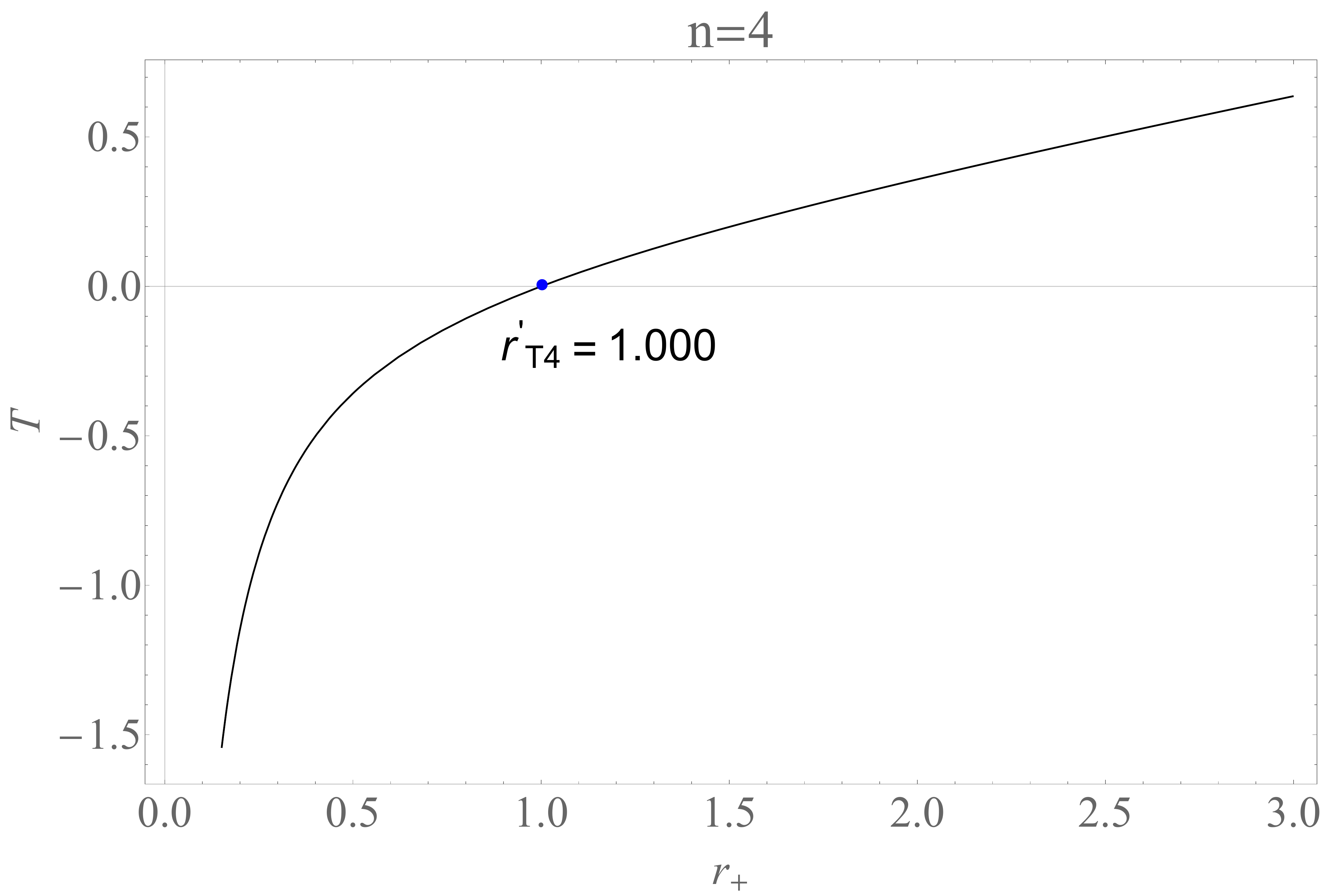}
        \includegraphics[angle=0,width=0.29\textwidth]{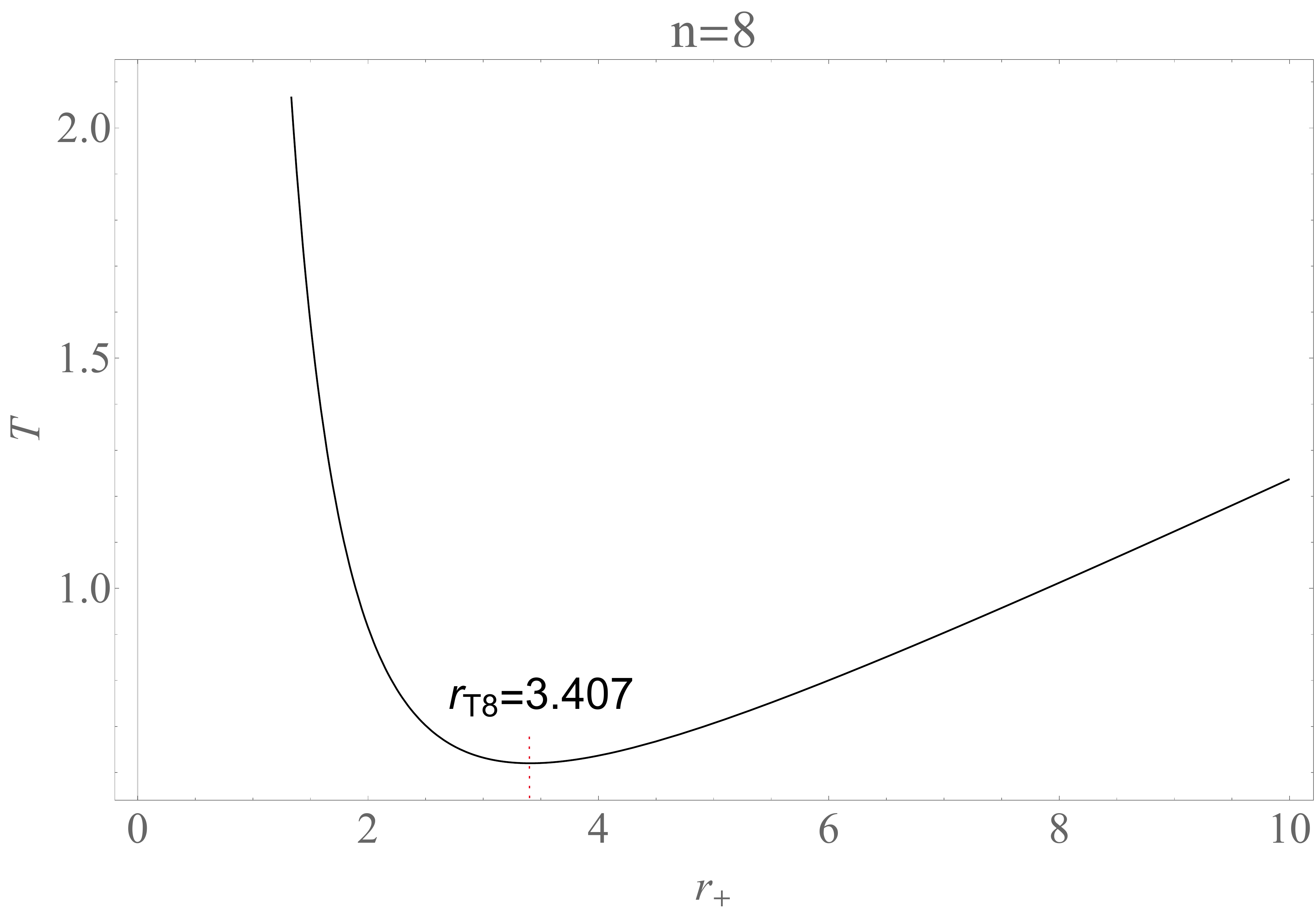}
	\caption{Temperature $T$ as a function of horizon radius $r_{+}$ for $R_{0}=-12, b=1.5.$ Left: $n=4, \Phi=1, r_{T_{4}}=0.430$. Middle: $n=4, \Phi=3,  r_{T_{4}}^{'}=1.000$, Right: $n=8, \Phi=3, r_{T_{8}}=3.407$.}
  \label{figt}
\end{figure}

 \begin{figure}[H]
 \centering
        \includegraphics[angle=0,width=0.295\textwidth]{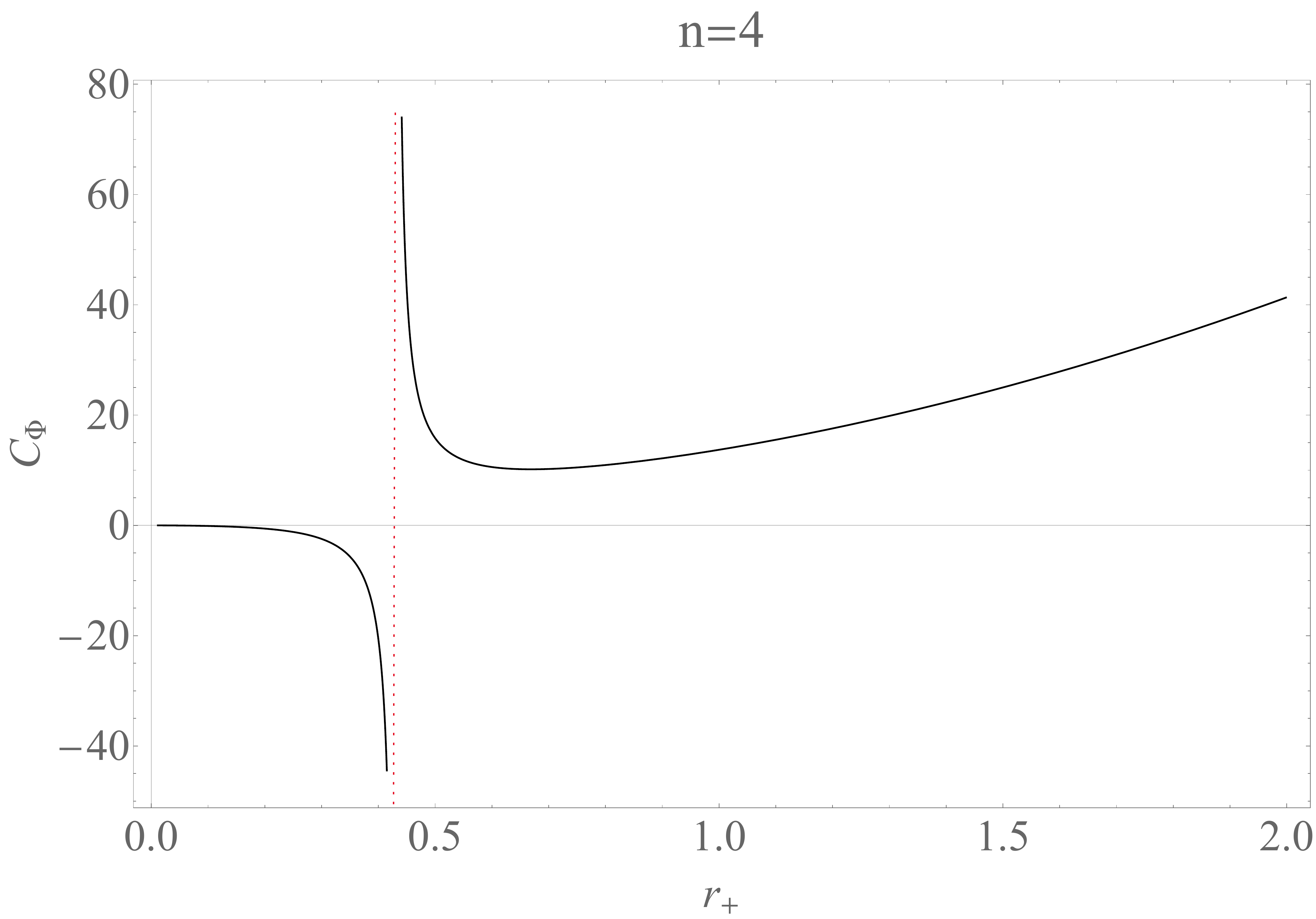}
       \includegraphics[angle=0,width=0.29\textwidth]{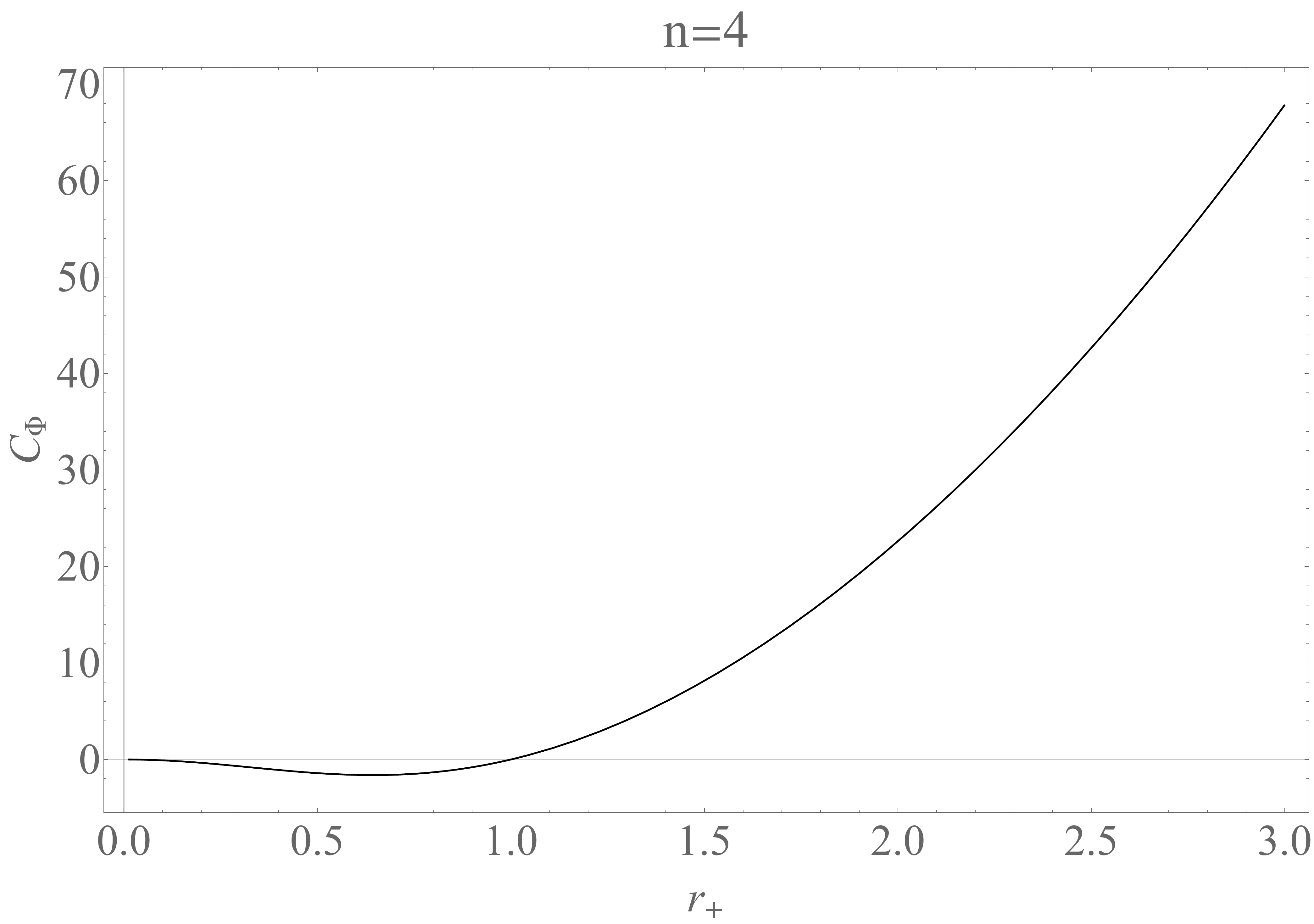}
         \includegraphics[angle=0,width=0.315\textwidth]{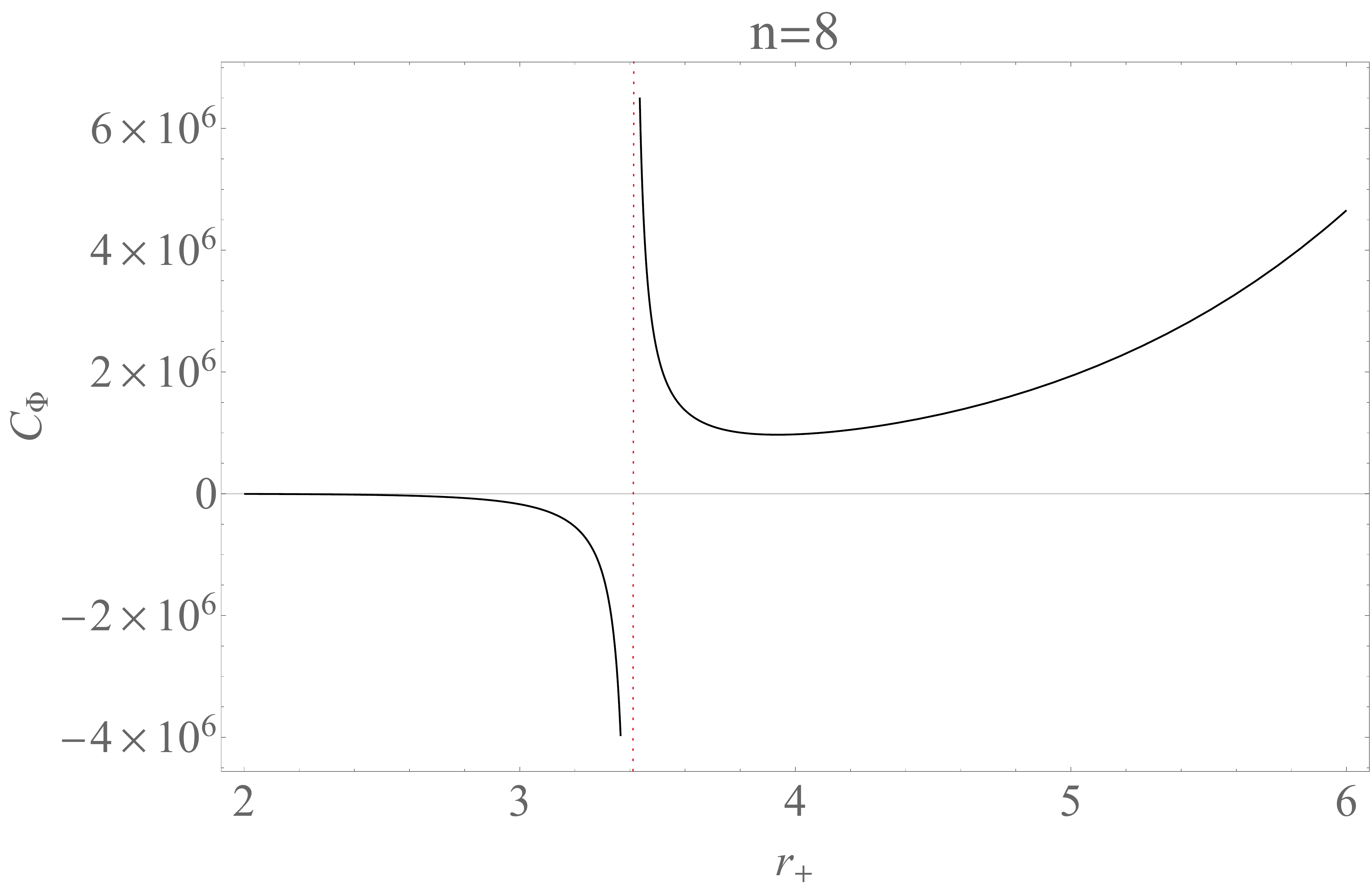}
	\caption{Specific heat capacity $C_{\Phi}$ as a function of horizon radius $r_{+}$ for $R_{0}=-12, b=1.5.$ Left: $n=4, \Phi=1$. Middle: $n=4, \Phi=3$. Right: $n=8, \Phi=3$.}
  \label{figc}
\end{figure}

 \begin{figure}[H]
 \centering
        \includegraphics[angle=0,width=0.29\textwidth]{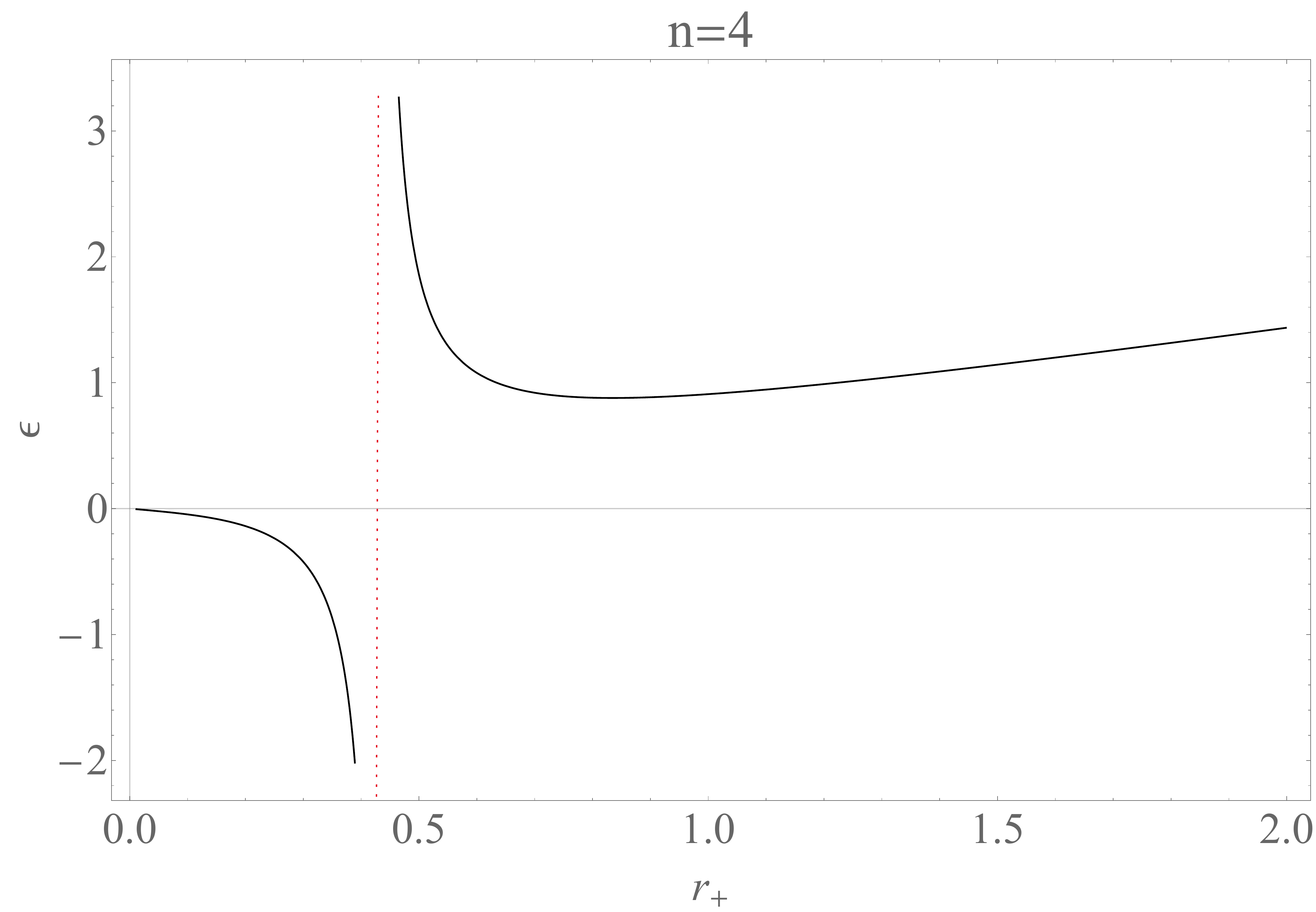}
       \includegraphics[angle=0,width=0.29\textwidth]{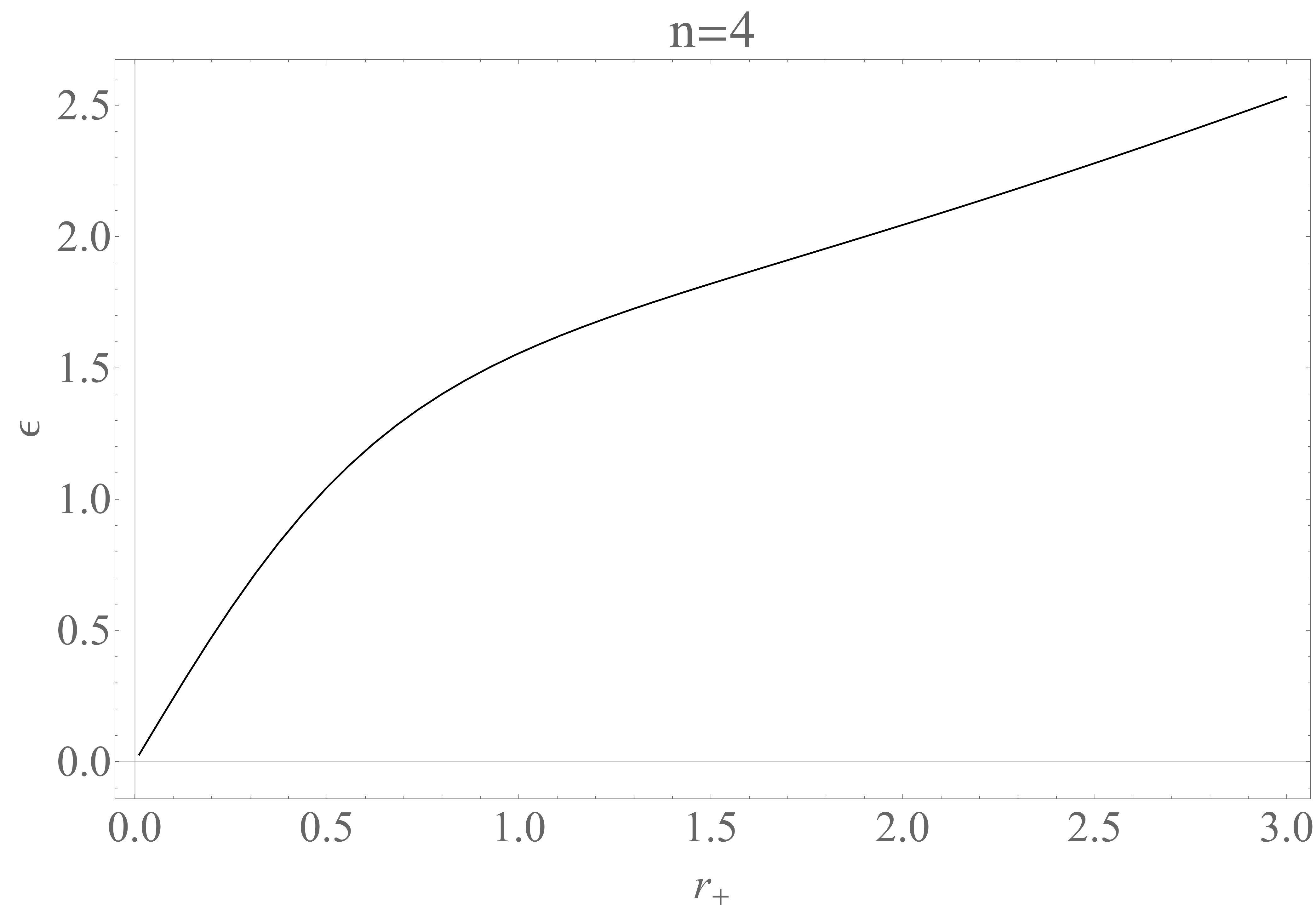}
         \includegraphics[angle=0,width=0.31\textwidth]{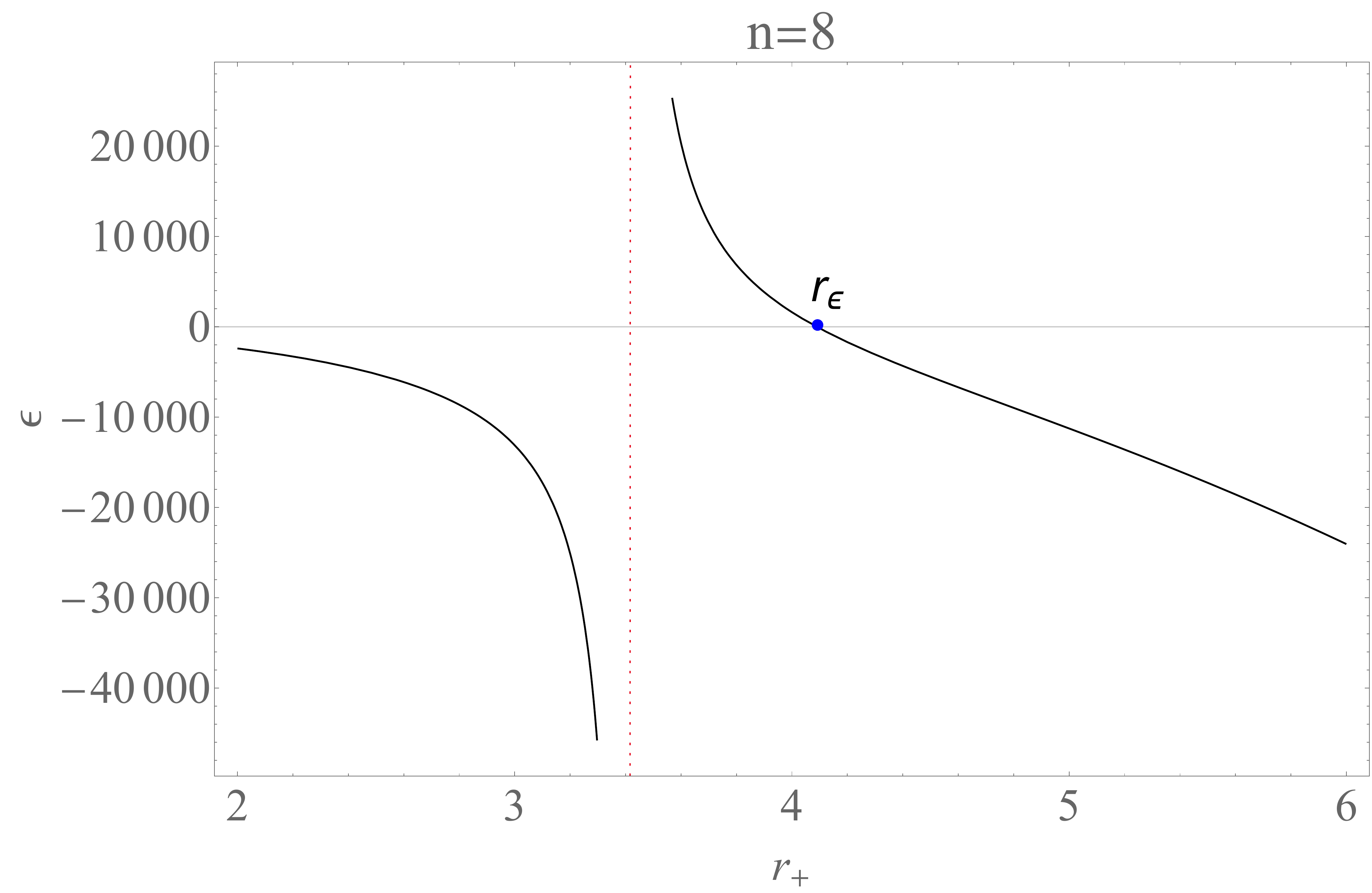}
	\caption{electrical permittivity $\epsilon$ as a function of horizon radius $r_{+}$ for $R_{0}=-12, b=1.5.$ Left: $n=4, \Phi=1$. Middle: $n=4, \Phi=3$. Right: $n=8, \Phi=3, r_{\epsilon}=4.090$.}
  \label{fige}
\end{figure}

\begin{figure}[H]
 \centering
        \includegraphics[angle=0,width=0.296\textwidth]{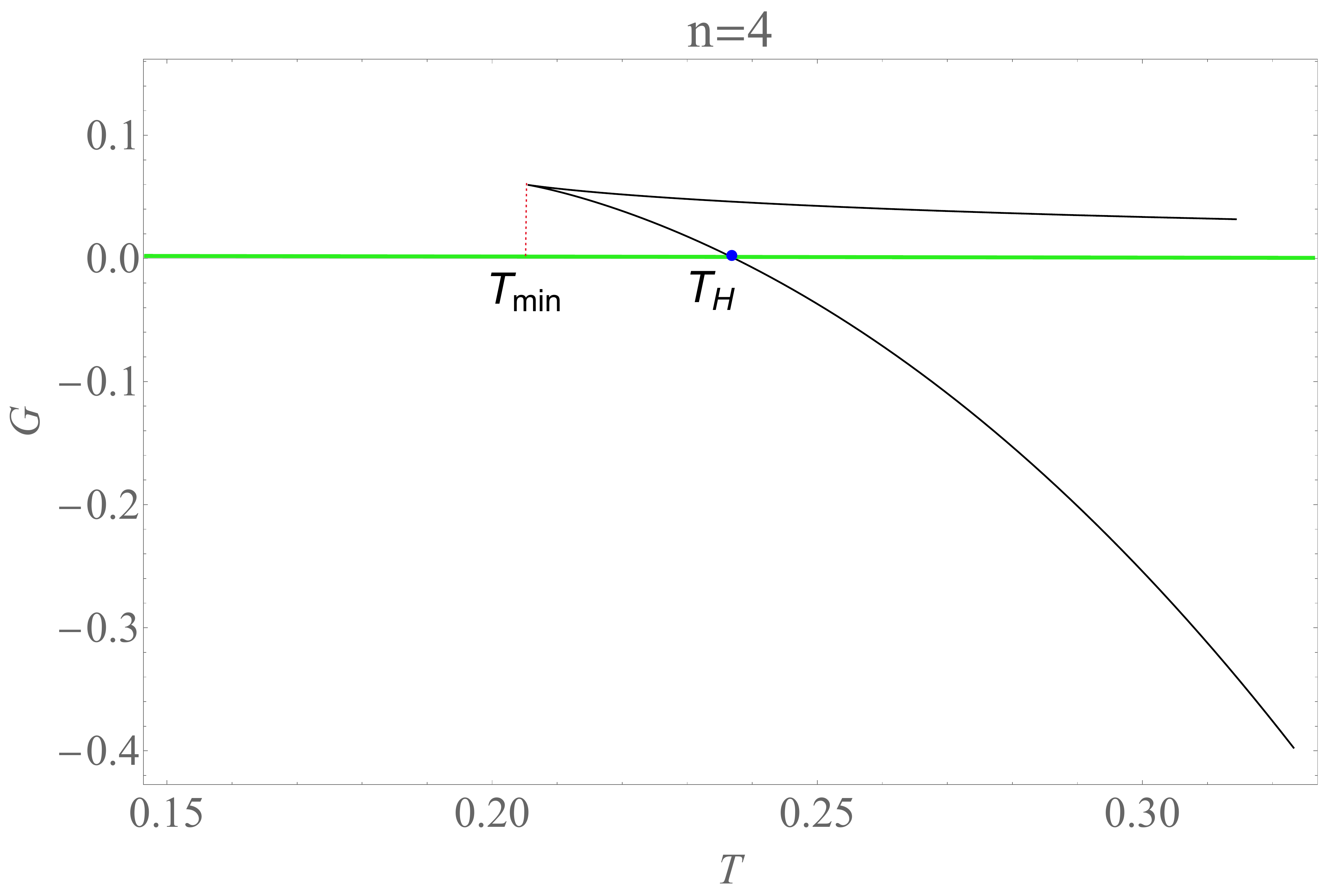}
       \includegraphics[angle=0,width=0.295\textwidth]{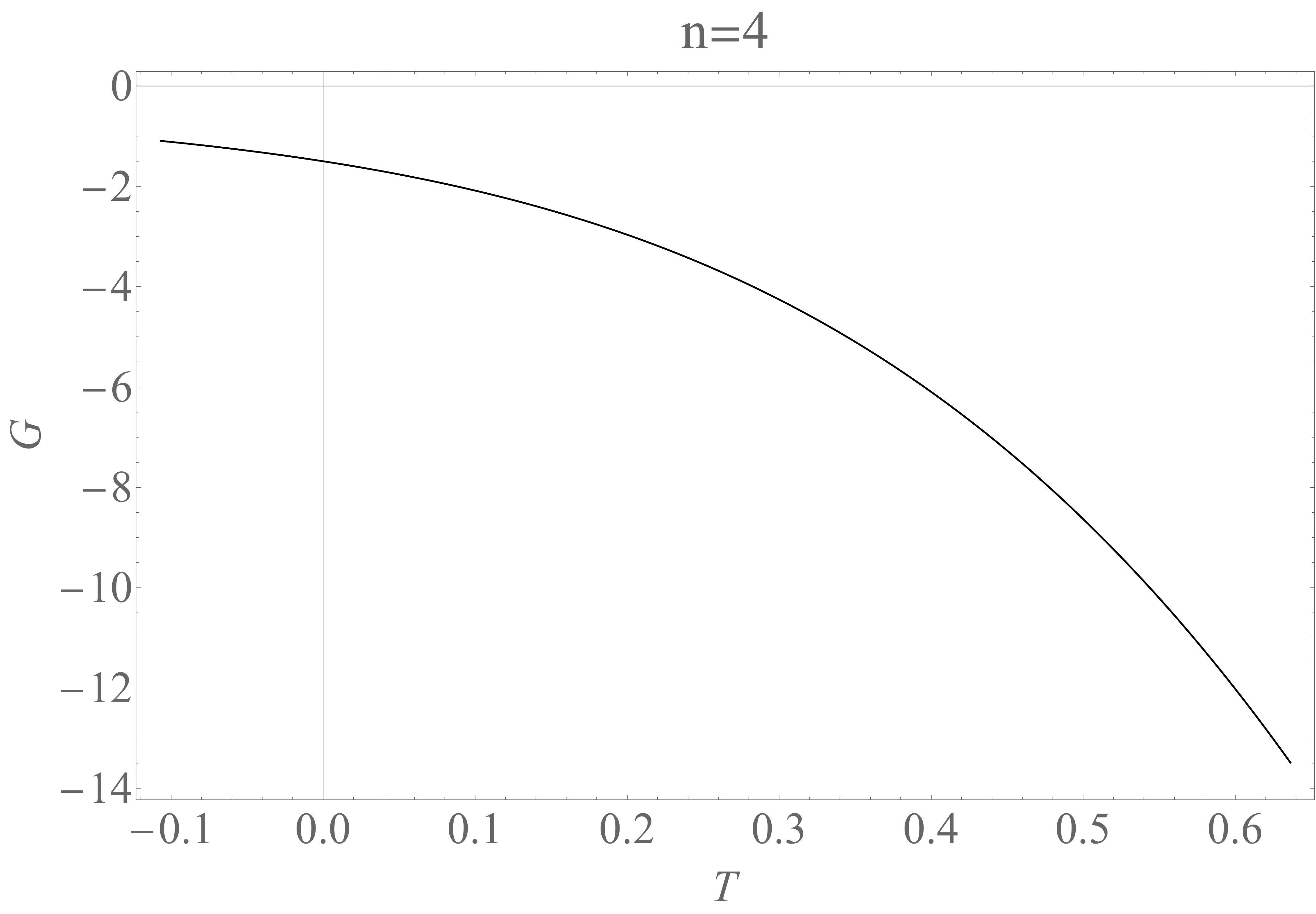}
         \includegraphics[angle=0,width=0.3\textwidth]{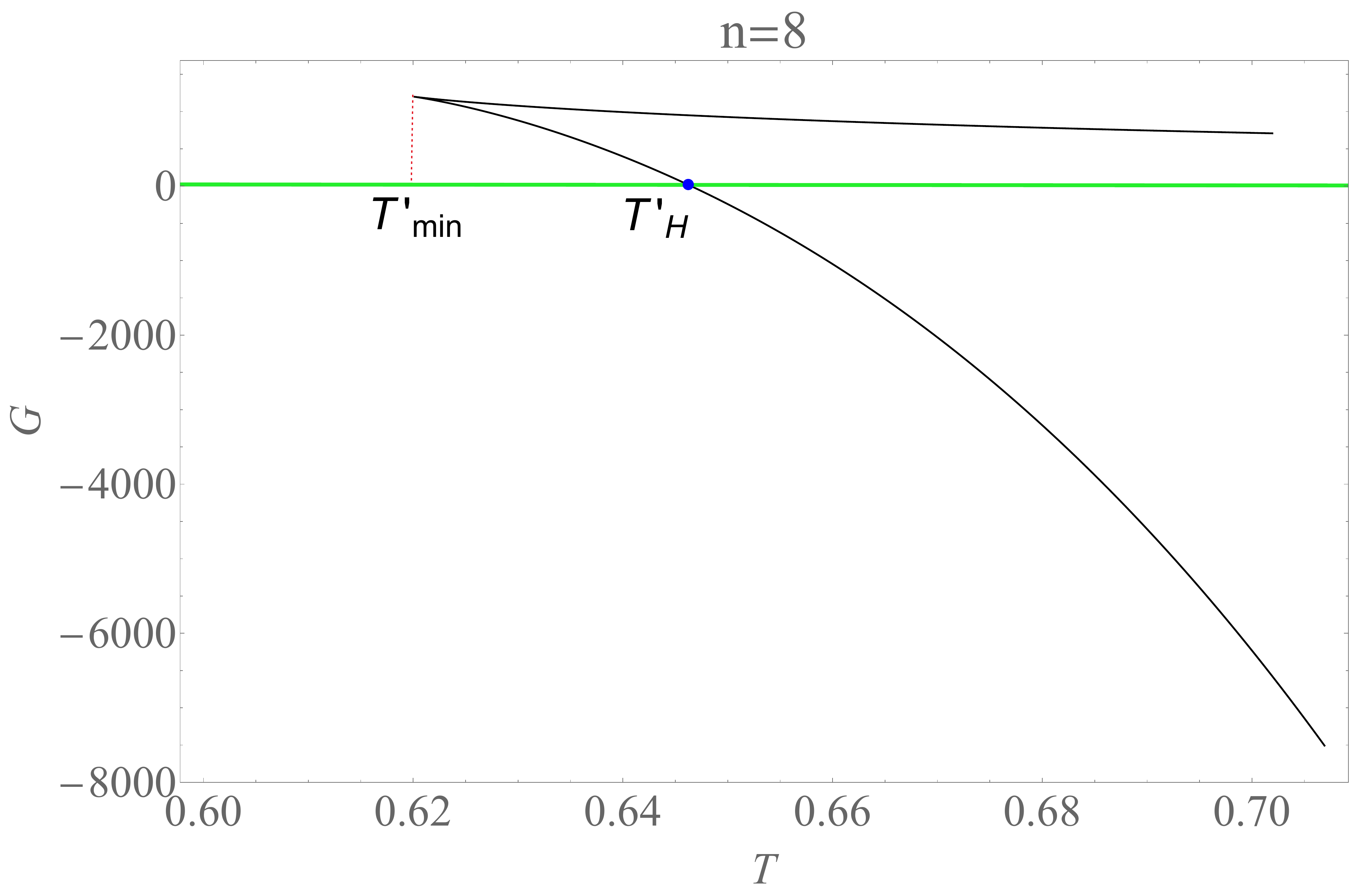}
	\caption{Gibbs free energy $G$ as a function of horizon radius $r_{+}$ for $R_{0}=-12, b=1.5.$ Left: $n=4, \Phi=1$. Middle: $n=4, \Phi=3$. Right: $n=8, \Phi=3$.}
  \label{figg}
\end{figure}

\begin{figure}[H]
 \centering
        \includegraphics[angle=0,width=0.3\textwidth]{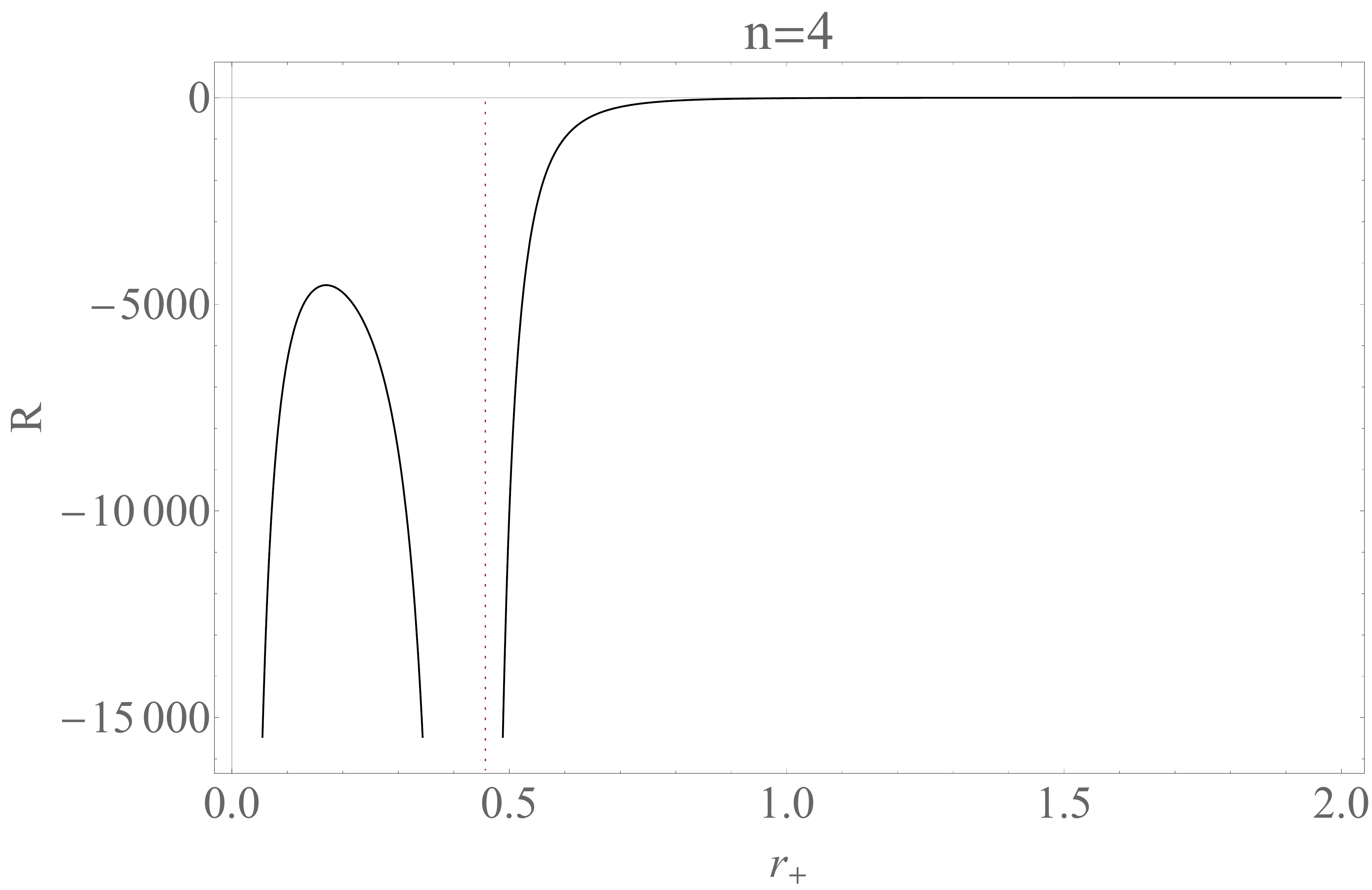}
       \includegraphics[angle=0,width=0.295\textwidth]{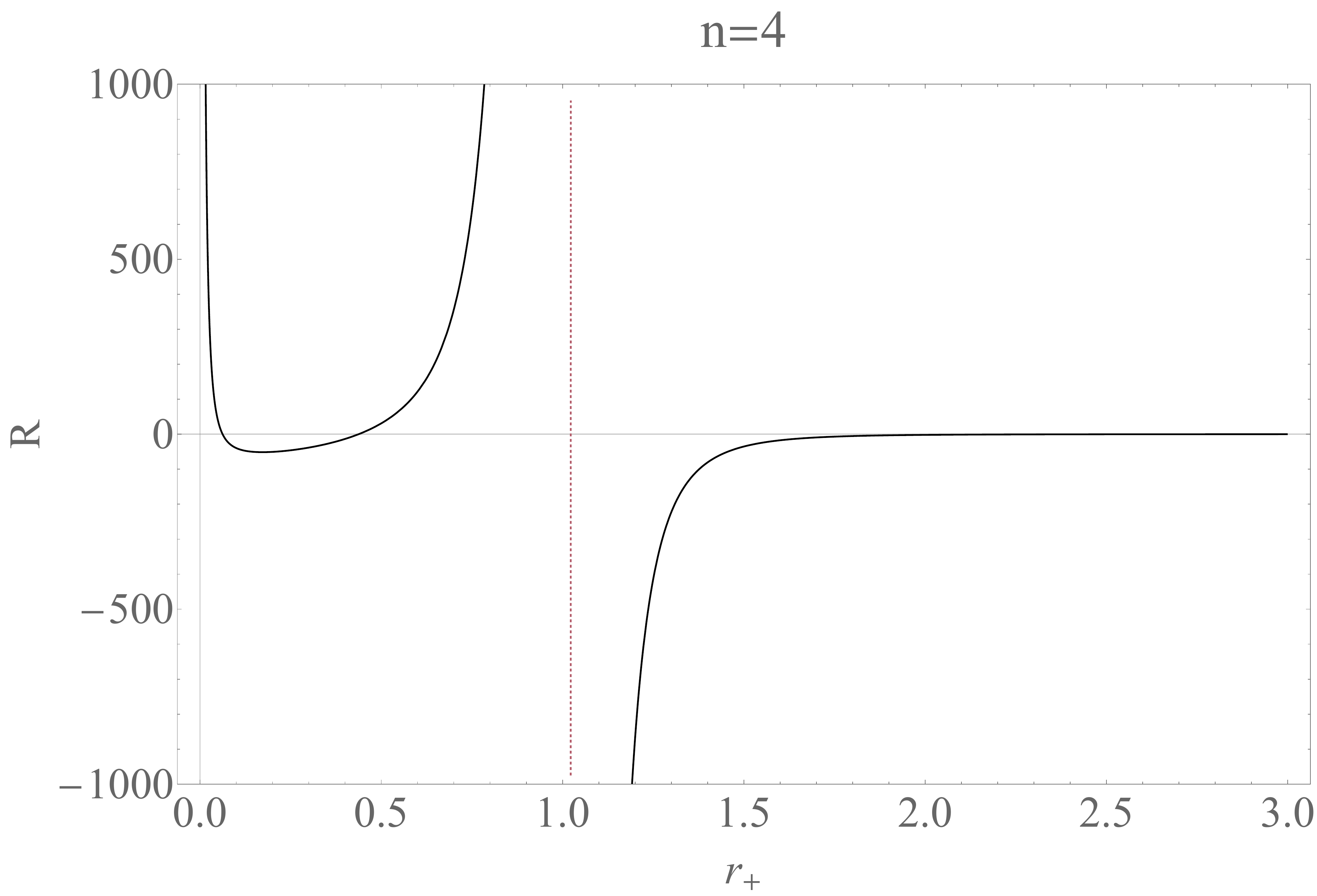}
           \includegraphics[angle=0,width=0.295\textwidth]{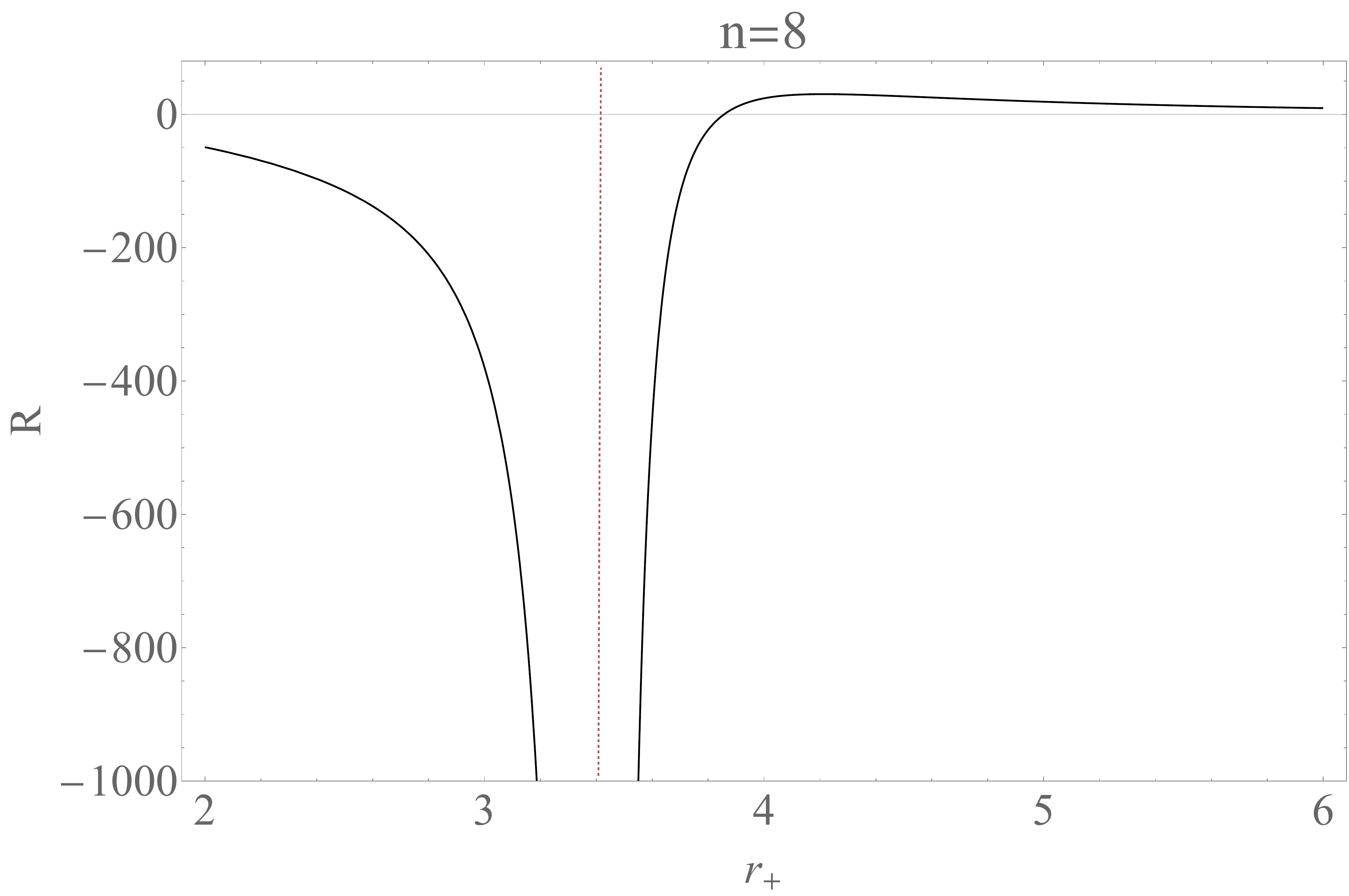}
	\caption{Intrinsic curvature $R$ as a function of horizon radius $r_{+}$ for $R_{0}=-12, b=1.5.$ Left: $n=4, \Phi=1$. Middle: $n=4, \Phi=3$. Right: $n=8, \Phi=3$.}
  \label{figr}
\end{figure}

\begin{figure}[H]
 \centering
        \includegraphics[angle=0,width=0.29\textwidth]{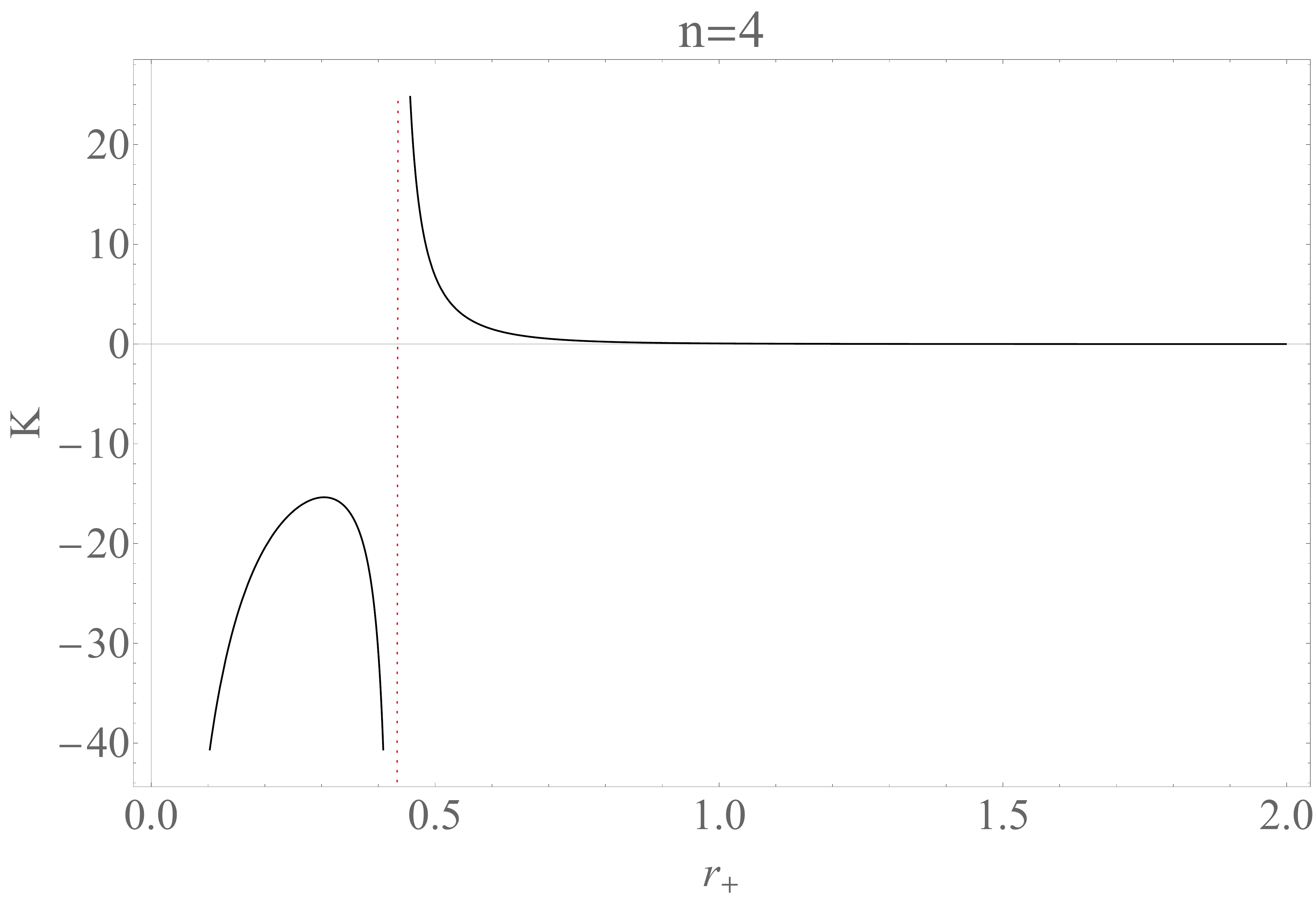}
       \includegraphics[angle=0,width=0.29\textwidth]{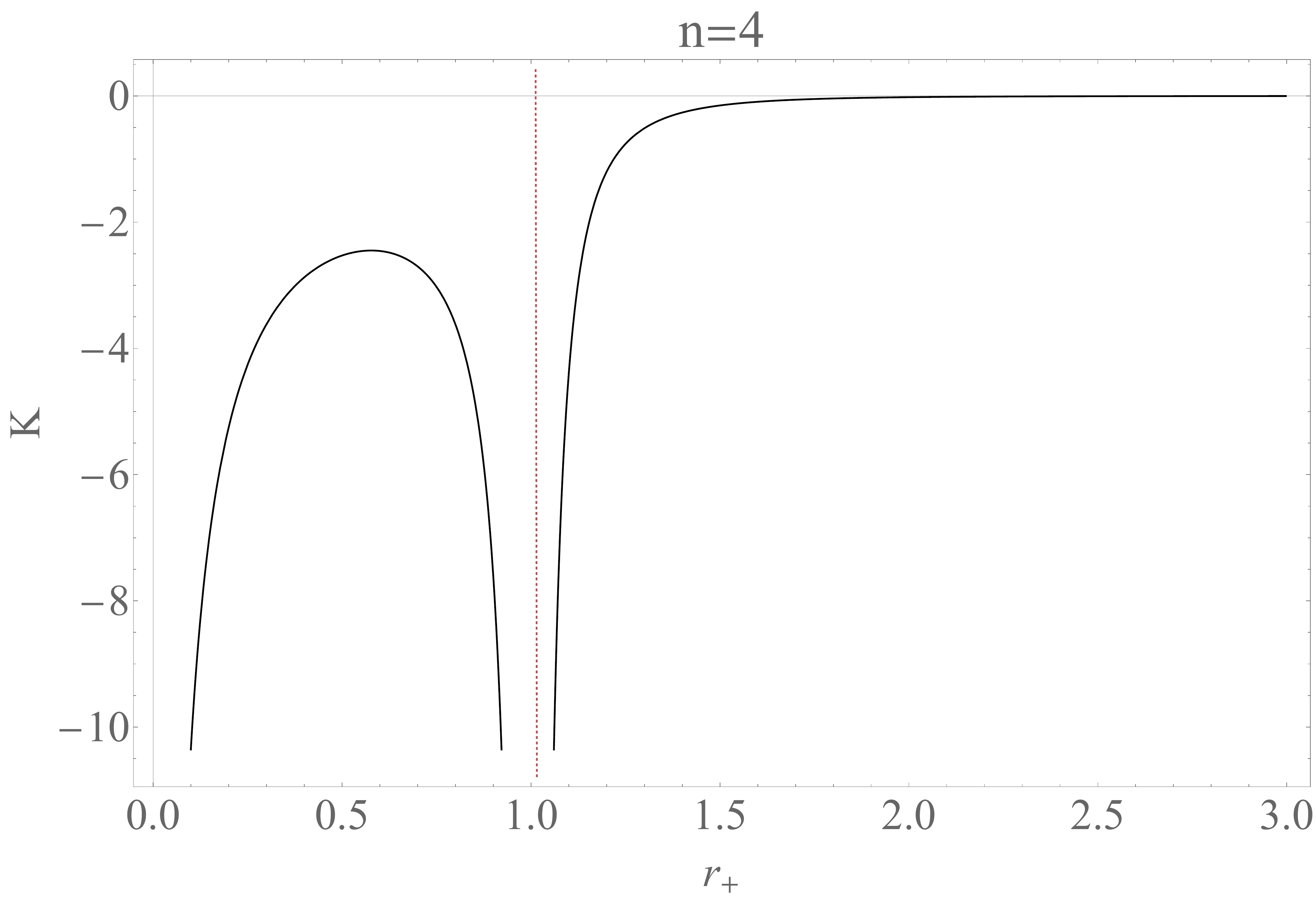}
         \includegraphics[angle=0,width=0.307\textwidth]{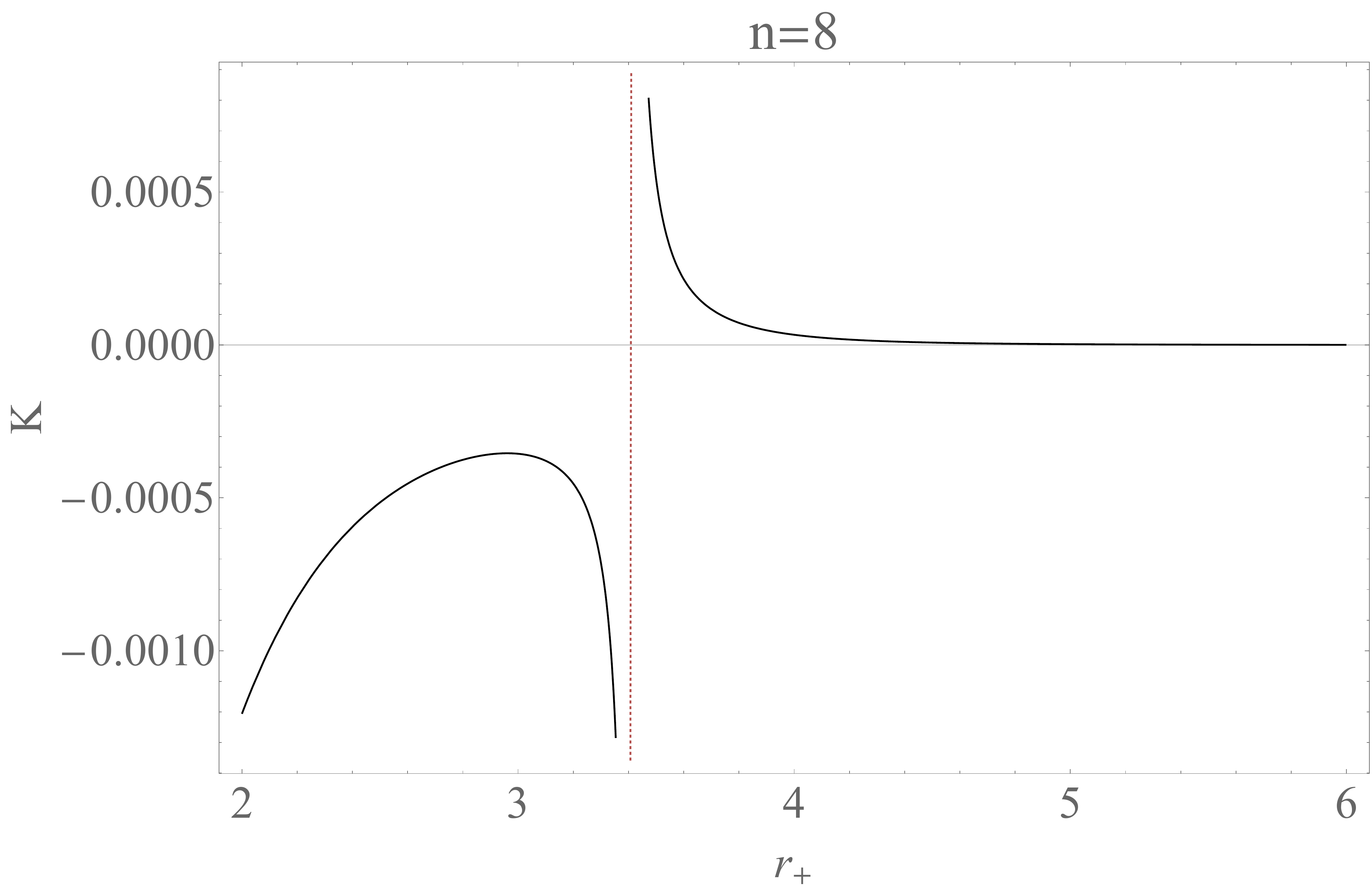}
	\caption{Extrinsic curvature $K$ as a function of horizon radius $r_{+}$ for $R_{0}=-12, b=1.5,$ Left:$n=4, \Phi=1$, Middle:$n=4, \Phi=3$, Right:$n=8, \Phi=3$.}
  \label{figk}
\end{figure}

\end{widetext}

\bibliographystyle{apsrev4-1}
\bibliography{Notes}

%merlin.mbs apsrev4-1.bst 2010-07-25 4.21a (PWD, AO, DPC) hacked
%Control: key (0)
%Control: author (72) initials jnrlst
%Control: editor formatted (1) identically to author
%Control: production of article title (-1) disabled
%Control: page (0) single
%Control: year (1) truncated
%Control: production of eprint (0) enabled
\begin{thebibliography}{32}%
\makeatletter
\providecommand \@ifxundefined [1]{%
 \@ifx{#1\undefined}
}%
\providecommand \@ifnum [1]{%
 \ifnum #1\expandafter \@firstoftwo
 \else \expandafter \@secondoftwo
 \fi
}%
\providecommand \@ifx [1]{%
 \ifx #1\expandafter \@firstoftwo
 \else \expandafter \@secondoftwo
 \fi
}%
\providecommand \natexlab [1]{#1}%
\providecommand \enquote  [1]{``#1''}%
\providecommand \bibnamefont  [1]{#1}%
\providecommand \bibfnamefont [1]{#1}%
\providecommand \citenamefont [1]{#1}%
\providecommand \href@noop [0]{\@secondoftwo}%
\providecommand \href [0]{\begingroup \@sanitize@url \@href}%
\providecommand \@href[1]{\@@startlink{#1}\@@href}%
\providecommand \@@href[1]{\endgroup#1\@@endlink}%
\providecommand \@sanitize@url [0]{\catcode `\\12\catcode `\$12\catcode
  `\&12\catcode `\#12\catcode `\^12\catcode `\_12\catcode `\%12\relax}%
\providecommand \@@startlink[1]{}%
\providecommand \@@endlink[0]{}%
\providecommand \url  [0]{\begingroup\@sanitize@url \@url }%
\providecommand \@url [1]{\endgroup\@href {#1}{\urlprefix }}%
\providecommand \urlprefix  [0]{URL }%
\providecommand \Eprint [0]{\href }%
\providecommand \doibase [0]{http://dx.doi.org/}%
\providecommand \selectlanguage [0]{\@gobble}%
\providecommand \bibinfo  [0]{\@secondoftwo}%
\providecommand \bibfield  [0]{\@secondoftwo}%
\providecommand \translation [1]{[#1]}%
\providecommand \BibitemOpen [0]{}%
\providecommand \bibitemStop [0]{}%
\providecommand \bibitemNoStop [0]{.\EOS\space}%
\providecommand \EOS [0]{\spacefactor3000\relax}%
\providecommand \BibitemShut  [1]{\csname bibitem#1\endcsname}%
\let\auto@bib@innerbib\@empty
%</preamble>
\bibitem [{\citenamefont {Maldacena}(1999)}]{Maldacena:1997re}%
  \BibitemOpen
  \bibfield  {author} {\bibinfo {author} {\bibfnamefont {J.~M.}\ \bibnamefont
  {Maldacena}},\ }\href {\doibase 10.1023/A:1026654312961} {\bibfield
  {journal} {\bibinfo  {journal} {Int. J. Theor. Phys.}\ }\textbf {\bibinfo
  {volume} {38}},\ \bibinfo {pages} {1113} (\bibinfo {year} {1999})},\ \bibinfo
  {note} {[Adv. Theor. Math. Phys.2,231(1998)]},\ \Eprint
  {http://arxiv.org/abs/hep-th/9711200} {arXiv:hep-th/9711200 [hep-th]}
  \BibitemShut {NoStop}%
%%CITATION = HEP-TH/9711200;%%
\bibitem [{\citenamefont {Witten}(1998{\natexlab{a}})}]{Witten:1998qj}%
  \BibitemOpen
  \bibfield  {author} {\bibinfo {author} {\bibfnamefont {E.}~\bibnamefont
  {Witten}},\ }\href@noop {} {\bibfield  {journal} {\bibinfo  {journal} {Adv.
  Theor. Math. Phys.}\ }\textbf {\bibinfo {volume} {2}},\ \bibinfo {pages}
  {253} (\bibinfo {year} {1998}{\natexlab{a}})},\ \Eprint
  {http://arxiv.org/abs/hep-th/9802150} {arXiv:hep-th/9802150 [hep-th]}
  \BibitemShut {NoStop}%
%%CITATION = HEP-TH/9802150;%%
\bibitem [{\citenamefont {Chamblin}\ \emph {et~al.}(1999)\citenamefont
  {Chamblin}, \citenamefont {Emparan}, \citenamefont {Johnson},\ and\
  \citenamefont {Myers}}]{Chamblin:1999tk}%
  \BibitemOpen
  \bibfield  {author} {\bibinfo {author} {\bibfnamefont {A.}~\bibnamefont
  {Chamblin}}, \bibinfo {author} {\bibfnamefont {R.}~\bibnamefont {Emparan}},
  \bibinfo {author} {\bibfnamefont {C.~V.}\ \bibnamefont {Johnson}}, \ and\
  \bibinfo {author} {\bibfnamefont {R.~C.}\ \bibnamefont {Myers}},\ }\href
  {\doibase 10.1103/PhysRevD.60.064018} {\bibfield  {journal} {\bibinfo
  {journal} {Phys. Rev.}\ }\textbf {\bibinfo {volume} {D60}},\ \bibinfo {pages}
  {064018} (\bibinfo {year} {1999})},\ \Eprint
  {http://arxiv.org/abs/hep-th/9902170} {arXiv:hep-th/9902170 [hep-th]}
  \BibitemShut {NoStop}%
%%CITATION = HEP-TH/9902170;%%
\bibitem [{\citenamefont {Kastor}\ \emph {et~al.}(2009)\citenamefont {Kastor},
  \citenamefont {Ray},\ and\ \citenamefont {Traschen}}]{Kastor:2009wy}%
  \BibitemOpen
  \bibfield  {author} {\bibinfo {author} {\bibfnamefont {D.}~\bibnamefont
  {Kastor}}, \bibinfo {author} {\bibfnamefont {S.}~\bibnamefont {Ray}}, \ and\
  \bibinfo {author} {\bibfnamefont {J.}~\bibnamefont {Traschen}},\ }\href
  {\doibase 10.1088/0264-9381/26/19/195011} {\bibfield  {journal} {\bibinfo
  {journal} {Class. Quant. Grav.}\ }\textbf {\bibinfo {volume} {26}},\ \bibinfo
  {pages} {195011} (\bibinfo {year} {2009})},\ \Eprint
  {http://arxiv.org/abs/0904.2765} {arXiv:0904.2765 [hep-th]} \BibitemShut
  {NoStop}%
%%CITATION = ARXIV:0904.2765;%%
\bibitem [{\citenamefont {Kubiznak}\ and\ \citenamefont
  {Mann}(2012)}]{Kubiznak:2012wp}%
  \BibitemOpen
  \bibfield  {author} {\bibinfo {author} {\bibfnamefont {D.}~\bibnamefont
  {Kubiznak}}\ and\ \bibinfo {author} {\bibfnamefont {R.~B.}\ \bibnamefont
  {Mann}},\ }\href {\doibase 10.1007/JHEP07(2012)033} {\bibfield  {journal}
  {\bibinfo  {journal} {JHEP}\ }\textbf {\bibinfo {volume} {07}},\ \bibinfo
  {pages} {033} (\bibinfo {year} {2012})},\ \Eprint
  {http://arxiv.org/abs/1205.0559} {arXiv:1205.0559 [hep-th]} \BibitemShut
  {NoStop}%
%%CITATION = ARXIV:1205.0559;%%
\bibitem [{\citenamefont {Johnson}(2014)}]{Johnson:2014yja}%
  \BibitemOpen
  \bibfield  {author} {\bibinfo {author} {\bibfnamefont {C.~V.}\ \bibnamefont
  {Johnson}},\ }\href {\doibase 10.1088/0264-9381/31/20/205002} {\bibfield
  {journal} {\bibinfo  {journal} {Class. Quant. Grav.}\ }\textbf {\bibinfo
  {volume} {31}},\ \bibinfo {pages} {205002} (\bibinfo {year} {2014})},\
  \Eprint {http://arxiv.org/abs/1404.5982} {arXiv:1404.5982 [hep-th]}
  \BibitemShut {NoStop}%
%%CITATION = ARXIV:1404.5982;%%
\bibitem [{\citenamefont {Wei}\ and\ \citenamefont {Liu}(2015)}]{Wei:2015iwa}%
  \BibitemOpen
  \bibfield  {author} {\bibinfo {author} {\bibfnamefont {S.-W.}\ \bibnamefont
  {Wei}}\ and\ \bibinfo {author} {\bibfnamefont {Y.-X.}\ \bibnamefont {Liu}},\
  }\href {\doibase 10.1103/PhysRevLett.116.169903,
  10.1103/PhysRevLett.115.111302} {\bibfield  {journal} {\bibinfo  {journal}
  {Phys. Rev. Lett.}\ }\textbf {\bibinfo {volume} {115}},\ \bibinfo {pages}
  {111302} (\bibinfo {year} {2015})},\ \bibinfo {note} {[Erratum: Phys. Rev.
  Lett.116,no.16,169903(2016)]},\ \Eprint {http://arxiv.org/abs/1502.00386}
  {arXiv:1502.00386 [gr-qc]} \BibitemShut {NoStop}%
%%CITATION = ARXIV:1502.00386;%%
\bibitem [{\citenamefont {Pourhassan}\ and\ \citenamefont
  {Faizal}(2015)}]{Pourhassan:2015cga}%
  \BibitemOpen
  \bibfield  {author} {\bibinfo {author} {\bibfnamefont {B.}~\bibnamefont
  {Pourhassan}}\ and\ \bibinfo {author} {\bibfnamefont {M.}~\bibnamefont
  {Faizal}},\ }\href {\doibase 10.1209/0295-5075/111/40006} {\bibfield
  {journal} {\bibinfo  {journal} {EPL}\ }\textbf {\bibinfo {volume} {111}},\
  \bibinfo {pages} {40006} (\bibinfo {year} {2015})},\ \Eprint
  {http://arxiv.org/abs/1503.07418} {arXiv:1503.07418 [gr-qc]} \BibitemShut
  {NoStop}%
%%CITATION = ARXIV:1503.07418;%%
\bibitem [{\citenamefont {Zhang}(2018)}]{Zhang:2018nep}%
  \BibitemOpen
  \bibfield  {author} {\bibinfo {author} {\bibfnamefont {M.}~\bibnamefont
  {Zhang}},\ }\href {\doibase 10.1016/j.nuclphysb.2018.08.010} {\bibfield
  {journal} {\bibinfo  {journal} {Nucl. Phys.}\ }\textbf {\bibinfo {volume}
  {B935}},\ \bibinfo {pages} {170} (\bibinfo {year} {2018})}\BibitemShut
  {NoStop}%
%%CITATION = NUPHA,B935,170;%%
\bibitem [{\citenamefont {Kubiznak}\ \emph {et~al.}(2017)\citenamefont
  {Kubiznak}, \citenamefont {Mann},\ and\ \citenamefont
  {Teo}}]{Kubiznak:2016qmn}%
  \BibitemOpen
  \bibfield  {author} {\bibinfo {author} {\bibfnamefont {D.}~\bibnamefont
  {Kubiznak}}, \bibinfo {author} {\bibfnamefont {R.~B.}\ \bibnamefont {Mann}},
  \ and\ \bibinfo {author} {\bibfnamefont {M.}~\bibnamefont {Teo}},\ }\href
  {\doibase 10.1088/1361-6382/aa5c69} {\bibfield  {journal} {\bibinfo
  {journal} {Class. Quant. Grav.}\ }\textbf {\bibinfo {volume} {34}},\ \bibinfo
  {pages} {063001} (\bibinfo {year} {2017})},\ \Eprint
  {http://arxiv.org/abs/1608.06147} {arXiv:1608.06147 [hep-th]} \BibitemShut
  {NoStop}%
%%CITATION = ARXIV:1608.06147;%%
\bibitem [{\citenamefont {Banerjee}\ \emph {et~al.}(2017)\citenamefont
  {Banerjee}, \citenamefont {Majhi},\ and\ \citenamefont
  {Samanta}}]{Banerjee:2016nse}%
  \BibitemOpen
  \bibfield  {author} {\bibinfo {author} {\bibfnamefont {R.}~\bibnamefont
  {Banerjee}}, \bibinfo {author} {\bibfnamefont {B.~R.}\ \bibnamefont {Majhi}},
  \ and\ \bibinfo {author} {\bibfnamefont {S.}~\bibnamefont {Samanta}},\ }\href
  {\doibase 10.1016/j.physletb.2017.01.040} {\bibfield  {journal} {\bibinfo
  {journal} {Phys. Lett.}\ }\textbf {\bibinfo {volume} {B767}},\ \bibinfo
  {pages} {25} (\bibinfo {year} {2017})},\ \Eprint
  {http://arxiv.org/abs/1611.06701} {arXiv:1611.06701 [gr-qc]} \BibitemShut
  {NoStop}%
%%CITATION = ARXIV:1611.06701;%%
\bibitem [{\citenamefont {Lee}(2017)}]{Lee:2017ero}%
  \BibitemOpen
  \bibfield  {author} {\bibinfo {author} {\bibfnamefont {C.~O.}\ \bibnamefont
  {Lee}},\ }\href {\doibase 10.1016/j.physletb.2017.07.013} {\bibfield
  {journal} {\bibinfo  {journal} {Phys. Lett.}\ }\textbf {\bibinfo {volume}
  {B772}},\ \bibinfo {pages} {471} (\bibinfo {year} {2017})},\ \Eprint
  {http://arxiv.org/abs/1705.09047} {arXiv:1705.09047 [gr-qc]} \BibitemShut
  {NoStop}%
%%CITATION = ARXIV:1705.09047;%%
\bibitem [{\citenamefont {Wei}\ and\ \citenamefont {Liu}(2018)}]{Wei:2017mwc}%
  \BibitemOpen
  \bibfield  {author} {\bibinfo {author} {\bibfnamefont {S.-W.}\ \bibnamefont
  {Wei}}\ and\ \bibinfo {author} {\bibfnamefont {Y.-X.}\ \bibnamefont {Liu}},\
  }\href {\doibase 10.1103/PhysRevD.97.104027} {\bibfield  {journal} {\bibinfo
  {journal} {Phys. Rev.}\ }\textbf {\bibinfo {volume} {D97}},\ \bibinfo {pages}
  {104027} (\bibinfo {year} {2018})},\ \Eprint
  {http://arxiv.org/abs/1711.01522} {arXiv:1711.01522 [gr-qc]} \BibitemShut
  {NoStop}%
%%CITATION = ARXIV:1711.01522;%%
\bibitem [{\citenamefont {Chabab}\ \emph {et~al.}(2018)\citenamefont {Chabab},
  \citenamefont {El~Moumni}, \citenamefont {Iraoui}, \citenamefont {Masmar},\
  and\ \citenamefont {Zhizeh}}]{Chabab:2018lzf}%
  \BibitemOpen
  \bibfield  {author} {\bibinfo {author} {\bibfnamefont {M.}~\bibnamefont
  {Chabab}}, \bibinfo {author} {\bibfnamefont {H.}~\bibnamefont {El~Moumni}},
  \bibinfo {author} {\bibfnamefont {S.}~\bibnamefont {Iraoui}}, \bibinfo
  {author} {\bibfnamefont {K.}~\bibnamefont {Masmar}}, \ and\ \bibinfo {author}
  {\bibfnamefont {S.}~\bibnamefont {Zhizeh}},\ }\href {\doibase
  10.1016/j.physletb.2018.04.014} {\bibfield  {journal} {\bibinfo  {journal}
  {Phys. Lett.}\ }\textbf {\bibinfo {volume} {B781}},\ \bibinfo {pages} {316}
  (\bibinfo {year} {2018})},\ \Eprint {http://arxiv.org/abs/1804.03960}
  {arXiv:1804.03960 [hep-th]} \BibitemShut {NoStop}%
%%CITATION = ARXIV:1804.03960;%%
\bibitem [{\citenamefont {Hawking}\ and\ \citenamefont
  {Page}(1983)}]{Hawking:1982dh}%
  \BibitemOpen
  \bibfield  {author} {\bibinfo {author} {\bibfnamefont {S.~W.}\ \bibnamefont
  {Hawking}}\ and\ \bibinfo {author} {\bibfnamefont {D.~N.}\ \bibnamefont
  {Page}},\ }\href {\doibase 10.1007/BF01208266} {\bibfield  {journal}
  {\bibinfo  {journal} {Commun. Math. Phys.}\ }\textbf {\bibinfo {volume}
  {87}},\ \bibinfo {pages} {577} (\bibinfo {year} {1983})}\BibitemShut
  {NoStop}%
%%CITATION = CMPHA,87,577;%%
\bibitem [{\citenamefont {Witten}(1998{\natexlab{b}})}]{Witten:1998zw}%
  \BibitemOpen
  \bibfield  {author} {\bibinfo {author} {\bibfnamefont {E.}~\bibnamefont
  {Witten}},\ }\href@noop {} {\bibfield  {journal} {\bibinfo  {journal} {Adv.
  Theor. Math. Phys.}\ }\textbf {\bibinfo {volume} {2}},\ \bibinfo {pages}
  {505} (\bibinfo {year} {1998}{\natexlab{b}})},\ \Eprint
  {http://arxiv.org/abs/hep-th/9803131} {arXiv:hep-th/9803131 [hep-th]}
  \BibitemShut {NoStop}%
%%CITATION = HEP-TH/9803131;%%
\bibitem [{\citenamefont {De~Felice}\ and\ \citenamefont
  {Tsujikawa}(2010)}]{DeFelice:2010aj}%
  \BibitemOpen
  \bibfield  {author} {\bibinfo {author} {\bibfnamefont {A.}~\bibnamefont
  {De~Felice}}\ and\ \bibinfo {author} {\bibfnamefont {S.}~\bibnamefont
  {Tsujikawa}},\ }\href {\doibase 10.12942/lrr-2010-3} {\bibfield  {journal}
  {\bibinfo  {journal} {Living Rev. Rel.}\ }\textbf {\bibinfo {volume} {13}},\
  \bibinfo {pages} {3} (\bibinfo {year} {2010})},\ \Eprint
  {http://arxiv.org/abs/1002.4928} {arXiv:1002.4928 [gr-qc]} \BibitemShut
  {NoStop}%
%%CITATION = ARXIV:1002.4928;%%
\bibitem [{\citenamefont {Starobinsky}(1980)}]{Starobinsky:1980te}%
  \BibitemOpen
  \bibfield  {author} {\bibinfo {author} {\bibfnamefont {A.~A.}\ \bibnamefont
  {Starobinsky}},\ }\href {\doibase 10.1016/0370-2693(80)90670-X} {\bibfield
  {journal} {\bibinfo  {journal} {Phys. Lett.}\ }\textbf {\bibinfo {volume}
  {91B}},\ \bibinfo {pages} {99} (\bibinfo {year} {1980})}\BibitemShut
  {NoStop}%
%%CITATION = PHLTA,91B,99;%%
\bibitem [{\citenamefont {Altamirano}\ \emph {et~al.}(2013)\citenamefont
  {Altamirano}, \citenamefont {Kubiznak},\ and\ \citenamefont
  {Mann}}]{Altamirano:2013ane}%
  \BibitemOpen
  \bibfield  {author} {\bibinfo {author} {\bibfnamefont {N.}~\bibnamefont
  {Altamirano}}, \bibinfo {author} {\bibfnamefont {D.}~\bibnamefont
  {Kubiznak}}, \ and\ \bibinfo {author} {\bibfnamefont {R.~B.}\ \bibnamefont
  {Mann}},\ }\href {\doibase 10.1103/PhysRevD.88.101502} {\bibfield  {journal}
  {\bibinfo  {journal} {Phys. Rev.}\ }\textbf {\bibinfo {volume} {D88}},\
  \bibinfo {pages} {101502} (\bibinfo {year} {2013})},\ \Eprint
  {http://arxiv.org/abs/1306.5756} {arXiv:1306.5756 [hep-th]} \BibitemShut
  {NoStop}%
%%CITATION = ARXIV:1306.5756;%%
\bibitem [{\citenamefont {Dolan}(2014)}]{Dolan:2013yca}%
  \BibitemOpen
  \bibfield  {author} {\bibinfo {author} {\bibfnamefont {B.~P.}\ \bibnamefont
  {Dolan}},\ }\href {\doibase 10.1088/0264-9381/31/19/199601,
  10.1088/0264-9381/31/13/135012} {\bibfield  {journal} {\bibinfo  {journal}
  {Class. Quant. Grav.}\ }\textbf {\bibinfo {volume} {31}},\ \bibinfo {pages}
  {135012} (\bibinfo {year} {2014})},\ \bibinfo {note} {[Erratum: Class. Quant.
  Grav.31,no.19,199601(2014)]},\ \Eprint {http://arxiv.org/abs/1312.6810}
  {arXiv:1312.6810 [gr-qc]} \BibitemShut {NoStop}%
%%CITATION = ARXIV:1312.6810;%%
\bibitem [{\citenamefont {Carter}\ and\ \citenamefont
  {Neupane}(2005)}]{Carter:2005uw}%
  \BibitemOpen
  \bibfield  {author} {\bibinfo {author} {\bibfnamefont {B.~M.~N.}\
  \bibnamefont {Carter}}\ and\ \bibinfo {author} {\bibfnamefont {I.~P.}\
  \bibnamefont {Neupane}},\ }\href {\doibase 10.1103/PhysRevD.72.043534}
  {\bibfield  {journal} {\bibinfo  {journal} {Phys. Rev.}\ }\textbf {\bibinfo
  {volume} {D72}},\ \bibinfo {pages} {043534} (\bibinfo {year} {2005})},\
  \Eprint {http://arxiv.org/abs/gr-qc/0506103} {arXiv:gr-qc/0506103 [gr-qc]}
  \BibitemShut {NoStop}%
%%CITATION = GR-QC/0506103;%%
\bibitem [{\citenamefont {Cai}\ \emph {et~al.}(2015)\citenamefont {Cai},
  \citenamefont {Hu}, \citenamefont {Pan},\ and\ \citenamefont
  {Zhang}}]{Cai:2014znn}%
  \BibitemOpen
  \bibfield  {author} {\bibinfo {author} {\bibfnamefont {R.-G.}\ \bibnamefont
  {Cai}}, \bibinfo {author} {\bibfnamefont {Y.-P.}\ \bibnamefont {Hu}},
  \bibinfo {author} {\bibfnamefont {Q.-Y.}\ \bibnamefont {Pan}}, \ and\
  \bibinfo {author} {\bibfnamefont {Y.-L.}\ \bibnamefont {Zhang}},\ }\href
  {\doibase 10.1103/PhysRevD.91.024032} {\bibfield  {journal} {\bibinfo
  {journal} {Phys. Rev.}\ }\textbf {\bibinfo {volume} {D91}},\ \bibinfo {pages}
  {024032} (\bibinfo {year} {2015})},\ \Eprint {http://arxiv.org/abs/1409.2369}
  {arXiv:1409.2369 [hep-th]} \BibitemShut {NoStop}%
%%CITATION = ARXIV:1409.2369;%%
\bibitem [{\citenamefont {Fernando}(2016{\natexlab{a}})}]{Fernando:2016sps}%
  \BibitemOpen
  \bibfield  {author} {\bibinfo {author} {\bibfnamefont {S.}~\bibnamefont
  {Fernando}},\ }\href {\doibase 10.1103/PhysRevD.94.124049} {\bibfield
  {journal} {\bibinfo  {journal} {Phys. Rev.}\ }\textbf {\bibinfo {volume}
  {D94}},\ \bibinfo {pages} {124049} (\bibinfo {year} {2016}{\natexlab{a}})},\
  \Eprint {http://arxiv.org/abs/1611.05329} {arXiv:1611.05329 [gr-qc]}
  \BibitemShut {NoStop}%
%%CITATION = ARXIV:1611.05329;%%
\bibitem [{\citenamefont {Fernando}(2016{\natexlab{b}})}]{Fernando:2016qhq}%
  \BibitemOpen
  \bibfield  {author} {\bibinfo {author} {\bibfnamefont {S.}~\bibnamefont
  {Fernando}},\ }\href {\doibase 10.1142/S0217732316500966} {\bibfield
  {journal} {\bibinfo  {journal} {Mod. Phys. Lett.}\ }\textbf {\bibinfo
  {volume} {A31}},\ \bibinfo {pages} {1650096} (\bibinfo {year}
  {2016}{\natexlab{b}})},\ \Eprint {http://arxiv.org/abs/1605.04860}
  {arXiv:1605.04860 [gr-qc]} \BibitemShut {NoStop}%
%%CITATION = ARXIV:1605.04860;%%
\bibitem [{\citenamefont {Li}\ and\ \citenamefont {Mo}(2016)}]{Li:2016wzx}%
  \BibitemOpen
  \bibfield  {author} {\bibinfo {author} {\bibfnamefont {G.-Q.}\ \bibnamefont
  {Li}}\ and\ \bibinfo {author} {\bibfnamefont {J.-X.}\ \bibnamefont {Mo}},\
  }\href {\doibase 10.1103/PhysRevD.93.124021} {\bibfield  {journal} {\bibinfo
  {journal} {Phys. Rev.}\ }\textbf {\bibinfo {volume} {D93}},\ \bibinfo {pages}
  {124021} (\bibinfo {year} {2016})},\ \Eprint
  {http://arxiv.org/abs/1605.09121} {arXiv:1605.09121 [gr-qc]} \BibitemShut
  {NoStop}%
%%CITATION = ARXIV:1605.09121;%%
\bibitem [{\citenamefont {Dehghani}(2017)}]{Dehghani:2017zkm}%
  \BibitemOpen
  \bibfield  {author} {\bibinfo {author} {\bibfnamefont {M.}~\bibnamefont
  {Dehghani}},\ }\href {\doibase 10.1103/PhysRevD.96.044014} {\bibfield
  {journal} {\bibinfo  {journal} {Phys. Rev.}\ }\textbf {\bibinfo {volume}
  {D96}},\ \bibinfo {pages} {044014} (\bibinfo {year} {2017})}\BibitemShut
  {NoStop}%
%%CITATION = PHRVA,D96,044014;%%
\bibitem [{\citenamefont {Dehghani}\ and\ \citenamefont
  {Hamidi}(2017)}]{Dehghani:2017ckb}%
  \BibitemOpen
  \bibfield  {author} {\bibinfo {author} {\bibfnamefont {M.}~\bibnamefont
  {Dehghani}}\ and\ \bibinfo {author} {\bibfnamefont {S.}~\bibnamefont
  {Hamidi}},\ }\href {\doibase 10.1103/PhysRevD.96.044025} {\bibfield
  {journal} {\bibinfo  {journal} {Phys. Rev.}\ }\textbf {\bibinfo {volume}
  {D96}},\ \bibinfo {pages} {044025} (\bibinfo {year} {2017})}\BibitemShut
  {NoStop}%
%%CITATION = PHRVA,D96,044025;%%
\bibitem [{\citenamefont {Sheykhi}(2012)}]{Sheykhi:2012zz}%
  \BibitemOpen
  \bibfield  {author} {\bibinfo {author} {\bibfnamefont {A.}~\bibnamefont
  {Sheykhi}},\ }\href {\doibase 10.1103/PhysRevD.86.024013} {\bibfield
  {journal} {\bibinfo  {journal} {Phys. Rev.}\ }\textbf {\bibinfo {volume}
  {D86}},\ \bibinfo {pages} {024013} (\bibinfo {year} {2012})},\ \Eprint
  {http://arxiv.org/abs/1209.2960} {arXiv:1209.2960 [hep-th]} \BibitemShut
  {NoStop}%
%%CITATION = ARXIV:1209.2960;%%
\bibitem [{\citenamefont {Quevedo}(2007)}]{Quevedo:2006xk}%
  \BibitemOpen
  \bibfield  {author} {\bibinfo {author} {\bibfnamefont {H.}~\bibnamefont
  {Quevedo}},\ }\href {\doibase 10.1063/1.2409524} {\bibfield  {journal}
  {\bibinfo  {journal} {J. Math. Phys.}\ }\textbf {\bibinfo {volume} {48}},\
  \bibinfo {pages} {013506} (\bibinfo {year} {2007})},\ \Eprint
  {http://arxiv.org/abs/physics/0604164} {arXiv:physics/0604164 [physics]}
  \BibitemShut {NoStop}%
%%CITATION = PHYSICS/0604164;%%
\bibitem [{\citenamefont {Mansoori}\ \emph {et~al.}(2016)\citenamefont
  {Mansoori}, \citenamefont {Mirza},\ and\ \citenamefont
  {Sharifian}}]{Mansoori:2016jer}%
  \BibitemOpen
  \bibfield  {author} {\bibinfo {author} {\bibfnamefont {S.~A.~H.}\
  \bibnamefont {Mansoori}}, \bibinfo {author} {\bibfnamefont {B.}~\bibnamefont
  {Mirza}}, \ and\ \bibinfo {author} {\bibfnamefont {E.}~\bibnamefont
  {Sharifian}},\ }\href {\doibase 10.1016/j.physletb.2016.05.096} {\bibfield
  {journal} {\bibinfo  {journal} {Phys. Lett.}\ }\textbf {\bibinfo {volume}
  {B759}},\ \bibinfo {pages} {298} (\bibinfo {year} {2016})},\ \Eprint
  {http://arxiv.org/abs/1602.03066} {arXiv:1602.03066 [gr-qc]} \BibitemShut
  {NoStop}%
%%CITATION = ARXIV:1602.03066;%%
\bibitem [{\citenamefont {Zhang}\ \emph {et~al.}(2018)\citenamefont {Zhang},
  \citenamefont {Wang},\ and\ \citenamefont {Liu}}]{Zhang:2018djl}%
  \BibitemOpen
  \bibfield  {author} {\bibinfo {author} {\bibfnamefont {M.}~\bibnamefont
  {Zhang}}, \bibinfo {author} {\bibfnamefont {X.-Y.}\ \bibnamefont {Wang}}, \
  and\ \bibinfo {author} {\bibfnamefont {W.-B.}\ \bibnamefont {Liu}},\ }\href
  {\doibase 10.1016/j.physletb.2018.06.061} {\bibfield  {journal} {\bibinfo
  {journal} {Phys. Lett.}\ }\textbf {\bibinfo {volume} {B783}},\ \bibinfo
  {pages} {169} (\bibinfo {year} {2018})}\BibitemShut {NoStop}%
%%CITATION = PHLTA,B783,169;%%
\bibitem [{\citenamefont {Wang}\ \emph {et~al.}(2018)\citenamefont {Wang},
  \citenamefont {Zhang},\ and\ \citenamefont {Liu}}]{Wang:2018civ}%
  \BibitemOpen
  \bibfield  {author} {\bibinfo {author} {\bibfnamefont {X.-Y.}\ \bibnamefont
  {Wang}}, \bibinfo {author} {\bibfnamefont {M.}~\bibnamefont {Zhang}}, \ and\
  \bibinfo {author} {\bibfnamefont {W.-B.}\ \bibnamefont {Liu}},\ }\href
  {\doibase 10.1140/epjc/s10052-018-6434-4} {\bibfield  {journal} {\bibinfo
  {journal} {Eur. Phys. J.}\ }\textbf {\bibinfo {volume} {C78}},\ \bibinfo
  {pages} {955} (\bibinfo {year} {2018})}\BibitemShut {NoStop}%
%%CITATION = EPHJA,C78,955;%%
\end{thebibliography}%


%merlin.mbs apsrev4-1.bst 2010-07-25 4.21a (PWD, AO, DPC) hacked
%Control: key (0)
%Control: author (72) initials jnrlst
%Control: editor formatted (1) identically to author
%Control: production of article title (-1) disabled
%Control: page (0) single
%Control: year (1) truncated
%Control: production of eprint (0) enabled
%

\end{document}